\documentclass[aps,pre,twocolumn,longbibliography,groupedaddress]{revtex4-1}
\usepackage{graphicx,amsmath,color}
\begin{document}

\title{Thermalized formulation of soft glassy rheology}
\author{Robert S. Hoy}
\email{rshoy@usf.edu}
\affiliation{Department of Physics, University of South Florida, Tampa, FL 33620, USA}
\date{\today}
\begin{abstract}
We present a version of soft glassy rheology that includes thermalized strain degrees of freedom.
It fully specifies systems' strain-history-dependent positions on their energy landscapes and therefore allows for quantitative analysis of their heterogeneous yielding dynamics and nonequilibrium deformation thermodynamics.
As a demonstration of the method, we illustrate the very different characteristics of fully-thermal and nearly-athermal plasticity by comparing results for thermalized and nonthermalized plastic flow.
\end{abstract}
\maketitle

\section{Introduction}
\label{sec:intro}

As far back as the work of Ree and Eyring \cite{ree55}, plastic deformation of solids has been modeled as being controlled by multiple relaxation processes with different characteristic rates.  
The energy landscape picture of Stillinger et.\ al.\ \cite{stillinger95,debenedetti01} allows it to be simultaneously viewed as being controlled by energy minima of broadly distributed depths and statistical weights.
Recently, a multitude of simulation studies have rather conclusively shown that structural glasses can be regarded as being composed of localized ``plastic zones'' with a wide range of thermodynamic and mechanical stabilities \cite{tsamados09,riggleman10b,manning11,rodney11,ding14,schoenholz14,swayamjyoti14,swayamjyoti16,patinet16}.
``Soft spots'' have smaller elastic moduli, lower activation energies, higher vibrational entropies, and yield first under deformation, while ``hard spots'' follow opposite trends \cite{tsamados09,riggleman10b,manning11,rodney11,ding14,schoenholz14,swayamjyoti14,swayamjyoti16,patinet16}.
Modern theories of plasticity such as soft glassy rheology (SGR) \cite{sollich97,sollich98,fielding00} and shear transformation zones (STZ) \cite{falk98,falk04} connect the energy-landscape and plastic-zone ideas, viewing amorphous solids as being composed of spatially localized plastic zones that directly correspond to basins in systems' energy landscapes with characteristic relaxation rates determined by the heights of their associated energy barriers.
Recent studies \cite{bouchbinder09a,bouchbinder09b,langer12,sollich12,bouchbinder13,fuereder13,kamrin14} have shown that the STZ and SGR theories are thermodynamically consistent and therefore amenable to rigorous nonequilibrium-thermodynamic treatment.
However, a particularly important open problem  \cite{bouchbinder09a,bouchbinder09b,langer12,sollich12,bouchbinder13,fuereder13,kamrin14} is determining the degree to which plastic flow is \textit{thermalized}, i.e.\ the degree to which the ``slow'' degrees of freedom corresponding to plastic zone configurations are in equilibrium with the ``fast'' degrees of freedom \cite{bouchbinder09a,bouchbinder09b} corresponding to localized motions of systems' constituent atoms and molecules.
Here we present a version of SGR theory that includes fully thermalized strain degrees of freedom and plastic flow.

Consider a system composed of plastic zones of activation energy $\mathcal{U}$.
Standard SGR theory, following the trap model \cite{bouchaud92,monthus96}, accounts for glassy systems' elastic heterogeneity by assuming these energies are exponentially distributed, i.e.\ by employing an exponential energy landscape $\rho(\mathcal{U}) = \tilde{\mathcal{U}}^{-1} \exp(-\mathcal{U}/\tilde{\mathcal{U}})$.
If the typical zone's activation energy $\tilde{\mathcal{U}} = \alpha k_B T_g$, then one can define the reduced variable $u = \mathcal{U}/k_B T_g$ and obtain the convenient form $\rho_\alpha(u) = \alpha^{-1}\exp(-u/\alpha)$.
Under an applied strain rate $\dot{\epsilon}$, these zones either deform elastically or yield.
Over a time interval $\Delta t$, zones yield with probability $1-\exp[-\Delta t/\tau(u,\epsilon^{el},x)]$, where 
\begin{equation}
\tau(u, \epsilon^{el},x) = \tau_0 \exp\left[\displaystyle\frac{u}{\alpha x}  \right] \exp\left[ -\displaystyle\frac{K_u (\epsilon^{el})^2}{2\alpha x} \right]
\label{eq:tausgr}
\end{equation}
is their characteristic relaxation time, $\epsilon^{el}$ is their elastic strain, $K_u$ is their dimensionless elastic modulus, and $x$ is the dimensionless ``noise temperature''.
The model exhibits a glass transition at $x = 1$ \cite{sollich97,sollich98}.
Note that while $\alpha$ has -- for convenience -- been set to unity in most published theoretical work \cite{sollich97,sollich98,fielding00,sollich12,bouchbinder13,fuereder13}, the distributions of plastic-zone activation energies in model glasses \cite{tsamados09,riggleman10b,manning11,rodney11,ding14,schoenholz14,swayamjyoti14,swayamjyoti16,patinet16} indicate $\tilde{\mathcal{U}} > k_B T_g$ and hence $\alpha > 1$ for many systems.

In SGR theory, when zones yield, they are removed (annihilated) and are typically replaced by new unstrained zones, again with values of $u$ drawn randomly from $\rho_\alpha(u)$.
However, as noted in the original papers \cite{sollich97,sollich98}, there is no physical reason to assume plastic zones are either initially unstrained or are replaced by new \textit{unstrained} zones upon yielding.
These assumptions are merely heuristics adopted for simplicity  that have been followed in most subsequent work \cite{sollich12,bouchbinder13,fuereder13,warren08,fielding14,merabia16,radhakrishnan17}.
Recent simulations \cite{schoenholz14,patinet16} have suggested that plastic zones often survive through multiple yielding events and hence are not always annihilated, but that their elastic strains ($\epsilon^{el}$) and spring constants ($\mathcal{K}$) do in general change upon yielding.
One simple way to treat such effects theoretically is to assume that zones are annihilated upon yielding, but that the newly created zones replacing them are drawn %randomly 
from an energy landscape $\rho^*(u,\epsilon^{el})$ that accounts for strain energy.
For the nearly athermal systems for which SGR was originally formulated (e.g.\ foams and pastes \cite{sollich97,sollich98}), it remains unclear how to construct such a landscape.
For \textit{thermal} systems such as metallic and polymeric glasses \cite{schuh07,roth16}, however, $\rho^*(u,\epsilon^{el})$ can be inferred from thermodynamics.
Here we adopt this approach, extending SGR to account for strain degrees of freedom in a thermodynamically consistent fashion and to treat thermalized plastic flow.
Our method's continuous formulation allows direct calculation of systems' nonequilibrium, strain-history-dependent positions on their energy landscapes, which in turn allows standard statistical mechanics to be employed for followup calculations.

\section{Thermalized version of SGR theory}
\label{sec:theory}

Following other recent work \cite{sollich12,bouchbinder13,fuereder13}, we formally treat amorphous materials (``systems'') as ensembles of plastic zones.
Material disorder is encoded in the functional form of $\rho(u)$.
There are no infinitely deep energy minima in a real glass.
For this reason (and for numerical convenience), we introduce a cutoff at $u_{max} = \alpha^2$, and impose it by multiplying $\rho_\alpha(u)$ by a cutoff function $C_\alpha(u) = 1-(u/\alpha^2)$.  
The use of such a cutoff function is supported by the extreme value statistics of low-energy states in disordered systems \cite{bouchaud97}.
Physically, $\mathcal{U}_{max} = \alpha^2 k_B T_g$ is the activation energy of the most stable plastic zone configurations that are compatible with the given system's microscopic interactions; recent soft-spot studies \cite{rodney11,swayamjyoti14,swayamjyoti16} suggest $\alpha^2 \simeq 10$.
The resulting zone depth distribution (i.e.\ the density of $u$-zones on the glass' energy landscape) is
\begin{equation}
\rho(u) = \displaystyle\frac{   \left[ 1 - (u/\alpha^2) \right] \exp( - u/\alpha)}{v_0 \left[\alpha -1 + \exp(-\alpha)\right]},
\label{eq:rho1}
\end{equation}
where $v_0$ is the typical volume of a plastic zone.
For simplicity (and following conventional SGR theory \cite{sollich97,sollich98,fielding00,sollich12,bouchbinder13,fuereder13}), we  assume that:\ (i) zones are structureless so that the strain-dependent density of states $\rho^*(u,\epsilon^{el})$ is a function only of $u$, i.e.\ $\rho^*(u, \epsilon^{el}) \equiv \rho(u)$; (ii) zone volumes are independent of $u$ and $\epsilon^{el}$.

Suppose that the occupation probability of zones with activation energy $\mathcal{U}$ and elastic strain $\epsilon^{el}$ is $p(u,\epsilon^{el})$.
The statistical weight of such zones is $w(u,\epsilon^{el}) = \rho(u) p(u,\epsilon^{el})$.
This construction is obviously amenable to thermodynamic treatment.
The average value of any material property $\zeta$ is given by 
\begin{equation}
\left<\zeta\right> = \displaystyle\int_0^{\alpha^2} \displaystyle\int_{-\infty}^\infty \zeta(u,\epsilon^{el}) w(u,\epsilon^{el}) d\epsilon^{el} du.
\label{eq:thermoavg}
\end{equation}
For arbitrary $x$, the thermodynamics of SGR-model systems are complicated, but still tractable \cite{sollich12,bouchbinder13,fuereder13}.
Here we will consider the simpler case where the typical energy scale $X = \alpha k_B T_g x$ associated with SGR-style ``noise'' is thermal in origin, i.e.\ $X = k_B T$.
Zones' relaxation time (inverse yielding rate; Eq.\ \ref{eq:tausgr}) therefore becomes
\begin{equation}
\tau(u, \epsilon^{el},T) = \tau_0 \exp\left[\displaystyle\frac{T_g}{T} u  \right] \exp\left[ -\displaystyle\frac{T_g }{T} \displaystyle\frac{K_u (\epsilon^{el})^2}{2} \right].
\label{eq:tautherm}
\end{equation}
Thus the mapping of $\tau$ from standard SGR \cite{sollich97,sollich98} to the present theory is quite simple: $x$ is just the ratio of the typical thermal energy $k_B T$ to the typical zone activation energy $\tilde{\mathcal{U}} = \alpha k_B T_g$.  See Table \ref{tab:mapping} for a further discussion of relations between the notation employed herein and that of Ref.\ \cite{sollich98}. 

We assume that systems' strain degrees of freedom are thermalized and therefore most zones have nonzero stress and strain even in undeformed systems.
The energy of a strained zone is $\mathcal{E}_u(\epsilon^{el}) = -\mathcal{U} + \mathcal{K}_u [\epsilon^{el}]^2/2$, where $\mathcal{K}_u = k_B T_g K_u$.
Systems' partition functions are given by
\begin{equation}
\mathcal{Z} = \displaystyle\int_0^{\alpha^2} \displaystyle\int_{-\infty}^\infty \rho(u) \exp[-\beta (\mathcal{E}_u(\epsilon^{el}) + \alpha^2 k_B T_g)] d\epsilon^{el} du, 
\label{eq:partition}
\end{equation}
where $\beta = (k_B T)^{-1}$.
In Eq.\ \ref{eq:partition}, zones with activation energy $\mathcal{U}$ and elastic strain $\epsilon^{el}$ have a Boltzmann factor $f_{Boltz}(u,\epsilon^{el},T) = \exp[-\beta (\mathcal{E}_u(\epsilon^{el}) + \alpha^2 k_B T_g)]$; their equilibrium occupation probability is $p_{eq}(u,\epsilon^{el},T) =  f_{Boltz}(u,\epsilon^{el})/\mathcal{Z}$.
Thus, in \textit{unstrained} systems, the equilibrium statistical weight of such zones is $w_{eq}(u,\epsilon^{el},T) = \rho(u) p_{eq}(u,\epsilon^{el},T)$.

Here we will consider an idealized, highly-aged initial condition wherein systems have reached thermal equilibrium, i.e.\ we assume the initial zone statistical weights are $w(u,\epsilon^{el}) = w_{eq}(u,\epsilon^{el},T)$.
Numerical tractability requires assuming that the maximum magnitude of the elastic strain $\epsilon^{el}$ in unstrained systems' $u$-zones is $\delta(u)$.
Then the thermalized initial condition becomes 
\small
\begin{equation}
w(u,\epsilon^{el}) = \bigg{\{ } 
\begin{array}{lll}
w_{eq}(u,\epsilon^{el},T) & , & 0 \leq u \leq \alpha^2\ \rm{and}\ |\epsilon^{el}| \leq \delta(u) \\
& & \\
0 & , & u > \alpha^2 \ \rm{or} \ |\epsilon^{el}| > \delta(u)
\end{array}.
\label{eq:ic1}
\end{equation}
\normalsize
Here and below, proper normalization of $w(u,\epsilon^{el})$ is maintained by replacing the exact partition function (Eq.\ \ref{eq:partition}) with $\mathcal{Z} = \int_0^{\alpha^2} \int_{-\delta(u)}^{\delta(u)} \rho(u) f_{Boltz}(u,\epsilon^{el})d\epsilon^{el} du$ and adjusting $w_{eq}(u,\epsilon^{el},T)$ accordingly.

One obvious choice for $\delta(u)$ is zones' zero-temperature yield strain $\epsilon^y_u$; states with larger $\epsilon^{el}$ are unstable at all temperatures \cite{sollich98}.
Here we adopt this choice.  
We define $K_u = 2 u k(u)$ so that plastic zones have spring constants $\mathcal{K}_u = 2\mathcal{U} k(u)$ and their zero-temperature yield strains are $\epsilon^y_u = 1/\sqrt{k(u)}$.
For simplicity, here we choose $k(u) = 400$ in order to give all zones the same value of $\epsilon^y_u$, specifically $\epsilon^y_u = .05$, a typical value for real metallic \cite{schuh07} and polymeric \cite{roth16} glasses.
Zones with low $u$ thus correspond to low-modulus soft spots \cite{riggleman10b,manning11,ding14,schoenholz14,patinet16}.
Note that other functional forms for $k(u)$ can be chosen to give other distributions of $\epsilon^y_u$ as desired.
For example, low-$u$ zones can be made to yield at smaller strains -- as is typical of soft spots  \cite{manning11,schoenholz14} -- by choosing $k(u) \propto u^{-1}$, which gives $\epsilon^y_u \propto \sqrt{u}$.

The two most common experimental deformation protocols are constant-strain-rate extension (or compression, or shear) and constant-applied-stress creep.
Here we will consider the former since it is conceptually simpler \cite{fielding00}.
We will discuss a scalar version of our theory, but all equations and results presented below are straightforwardly generalizable to tensorial stresses and strains using methods like those described in Refs.\ \cite{cates04,fuereder13}.
The macrosopic strain applied to the system is $\epsilon = \dot{\epsilon}t$.
Then the total configurational energy density $E(\epsilon)$ of strained systems is \cite{fuereder13}
\begin{equation}
\displaystyle\frac{E(\epsilon)}{k_B T_g} =  \displaystyle\int_0^{\alpha^2} \displaystyle\int_{-\delta(u)}^{\delta(u) + \epsilon} \left[\alpha^4 k(u) (\epsilon^{el})^2  - 1 \right] u w(u,\epsilon^{el}) d\epsilon^{el} du.
\label{eq:E1}
\end{equation}

In SGR theory, zones are structureless and have no internal entropy \cite{sollich98}.
From the statistical definition of entropy $S = -k_B \left< \ln(p) \right>$, strained systems' configurational entropy density is given by \cite{sollich12,fuereder13}
\begin{equation} 
\displaystyle\frac{S(\epsilon)}{k_B} = - \displaystyle\int_0^{\alpha^2} \displaystyle\int_{-\delta(u)}^{\delta(u) + \epsilon}  \ln\left[ p(u,\epsilon^{el}) \right] w(u,\epsilon^{el}) d\epsilon^{el} du.
\label{eq:S1}
\end{equation}
Since $X = k_BT$, systems' Helmholtz free energy density $F(\epsilon)$ satisfies the usual relation $F(\epsilon) = E(\epsilon) - TS(\epsilon)$. 
Note that this definition of entropy is chosen to give $S(\epsilon) = k_B \ln[\Omega(\epsilon)]$ in the $T \to \infty$ limit, where $\Omega(\epsilon) = (\Delta u \Delta \epsilon)^{-1} \int_{0}^{\alpha^2} \int_{-\delta(u)}^{\delta(u) + \epsilon} \rho(u) d\epsilon^{el} du$ is the volume and $\Delta u$ and $\Delta \epsilon$ are the ``quanta'' of phase space.
In any discretized calculation, in the limit of small $\Delta u$ and $\Delta\epsilon$, $S(0) = S_0 - b(T) \ln(\Delta x \Delta\epsilon)$, where $S_0$ is a reference value and $b(T)$ can be determined by comparing unstrained systems with different $\Delta u \Delta \epsilon$ [with $b(0) = 0$ and $b(\infty) = 1$].
Determining ``natural'' values of $\Delta u$ and $\Delta \epsilon$ would require specifying the distinguishability of basins of different $u$ and $\epsilon^{el}$, which is beyond our scope; here we choose values of $\Delta u$ and $\Delta \epsilon$ that give clearly-converged results for $S(0) + \ln(\Delta u \Delta\epsilon)$ in the high-$T$ limit.

We evolve systems forward in time using the following plastic flow rule:
\begin{equation}
\begin{array}{l}
\displaystyle\frac{d w(u,\epsilon^{el})}{d t} = -\dot\epsilon\displaystyle\frac{\partial (u,\epsilon^{el})}{\partial \epsilon^{el}}   - \displaystyle\frac{ w(u,\epsilon^{el})}{\tau(u,\epsilon^{el},T)}  \\
\\
\ \ \ +\ f(u,\epsilon^{el},T) \displaystyle\int_0^{\alpha^2} \displaystyle\int_{-\delta(u)}^{\delta(u) + \epsilon}  \displaystyle\frac{w(\tilde{u},\tilde{\epsilon}^{el})}{\tau(\tilde{u},\tilde{\epsilon}^{el},T)} d\tilde{\epsilon}^{el} d\tilde{u}.
\end{array}
\label{eq:trapflow}
\end{equation}
Here $\tau^{-1}(u,\epsilon^{el},T)$ is the yielding rate of $u$-zones (Eq.\ \ref{eq:tautherm}), and the factor 
$f(u,\epsilon^{el},T)$ is given by
\begin{equation}
f(u,\epsilon^{el},T) = \displaystyle\frac{\rho(u) p_{eq}(u,\epsilon^{el},T) \theta(u,\epsilon^{el})} {\displaystyle\int_0^{\alpha^2} \displaystyle\int_{-\delta(u)}^{\delta(u)} \rho(u) p_{eq}(u,\epsilon^{el},T) d\tilde{\epsilon}^{el} d\tilde{u}},
\label{eq:fofx}
\end{equation}
where
\begin{equation}
\theta(u,\epsilon^{el}) = \bigg{\{} \begin{array}{ccl}
1 & , & |\epsilon^{el}| < \delta(u), \\
\\
0 & , &  |\epsilon^{el}| \geq \delta(u) 
\end{array}.
\label{eq:bounds}
\end{equation}
This form of $f(u,\epsilon^{el},T)$ ensures that newly created zones populate \textit{stable} [$\mathcal{E}_u(\epsilon^{el}) < 0$] configurations according to their equilibrium occupation probabilities.
A similar $f(u,\epsilon^{el},x)$ was proposed in Ref.\ \cite{sollich98} and was used to calculate the linear viscoelastic moduli $G^*(\omega)$; here we extend this method to nonlinear response.

The factors of $p_{eq}(u,\epsilon^{el},T)$ in Eq.\ \ref{eq:fofx} reflect the fact that the present theory is fundamentally thermal in nature, and is designed to treat \textit{thermalized} plastic deformation.
More specifically, the inclusion of the $p_{eq}(u,\epsilon^{el},T)$ terms reflects our assumption that plastic flow is thermalized by the same reservoir that maintains constant $T$.
Since we assume fully thermalized flow, we need not and do not adopt a dual-subsystem, two-temperature nonequilibrium-thermodynamic ansatz like those employed in Refs.\ 
\cite{bouchbinder09a,bouchbinder09b,sollich12,langer12,bouchbinder13,fuereder13,kamrin14}; cf.\ Section \ref{sec:conclude}.
Note that Ref.\ \cite{merabia16} similarly employed $x = k_BT/\tilde{\mathcal{U}}$ to treat plastic deformation of thermal glasses, but did not adopt a fully thermalized plastic flow rule including thermalized strain degrees of freedom as we have done here.

Two other technical points relating to differences between our theory and standard SGR should be mentioned.
First, in real systems, ``frustration'' \cite{sollich98} effects arising from correlations between spatially neighboring plastic zones may inhibit creation of new zones for which the sign of $\epsilon^{el}$ is opposite that of $\epsilon$.
Strong frustration would make a nonsymmetric $f(u,\epsilon^{el},T)$ more appropriate for describing plastic flow.
However, since proper treatments of frustration are presumably both complicated and system-specific \cite{sollich98}, they have rarely been treated within SGR theory, and are not considered here.
Second, we do not allow for the (very real \cite{schoenholz14,patinet16}) possibility that the number of plastic zones in a system changes during deformation, because any such changes are likely to be highly system-specific and thus beyond the scope of the present effort.

\begin{table}
\caption{Comparison of the notation used in this work to the notation used in Ref.\ \cite{sollich98}.  Note that the present model is mathematically equivalent to that discussed in Section IVC of Ref.\ \cite{sollich98} if one sets $\rho(E) = (1-E/\alpha)\exp(-E)$, $k(E)=2E$, and $q(\ell; E) = \rho(E) \exp[-k(E)\ell^2/2x]\theta(E,\ell)$.}
\begin{ruledtabular}
\begin{tabular}{lccc}
Quantity &  Ref.\ \cite{sollich98} & Present work \\
Scaled activation energy & $E$ & $u/\alpha$ \\
Scaled elastic strain & $\ell$ & $\epsilon^{el}/\epsilon^y_u$ \\
Flow factor & $q(\ell; E)$ & $f(u,\epsilon^{el},T)$ \\
Scaled noise temperature & $x$ & $k_B T/\tilde{\mathcal{U}}$ 
\end{tabular}
\end{ruledtabular}
\label{tab:mapping}
\end{table}

Eq.\ \ref{eq:trapflow} usually cannot be solved analytically, so we solve it numerically.
To make the model computationally tractable, we discretize the zone activation energies ($u_i = i\Delta u$) and strains ($\epsilon_j = j\Delta\epsilon$). 
This yields the evolution equation 
\begin{equation}
\begin{array}{c}
w(u_i, \epsilon_j; t_k) =  w(u_i, \epsilon_{j-1}; t_{k-1}) \left[1 -\displaystyle\frac{\Delta t}{\tau(u_i,\epsilon_{j-1},T)}\right] \\
\\
+\   f(u_i, \epsilon_j, T; t_{k-1}) \left< \tau^{-1} (\epsilon) \right> \Delta t,
\end{array}
\label{eq:flow}
\end{equation}
where the timestep $\Delta t = t_{k+1} - t_k = \Delta\epsilon/\dot\epsilon$ \cite{footexp}, 
and
\begin{equation}
\left< \tau^{-1} (\epsilon) \right> =  \displaystyle\int_0^{\alpha^2} \displaystyle\int_{-\delta(u)}^{\delta(u) + \epsilon} w(u,\epsilon^{el}) \tau^{-1}(u,\epsilon^{el},T) d\epsilon^{el}.
\label{eq:tauinv}
\end{equation}
is the average zone yielding rate.
On the right hand side of Eq.\ \ref{eq:flow}, the first term indicates zones present at the previous timestep ($t = t_{k-1}$) that did not yield, and the second term indicates creation of new zones with thermalized strains.
Since zones can yield at any time, the allowed values of $j$ at time $t_k$ are ($-\ell_i,-\ell_i+1,...,\ell_i + k$), where $\ell_i \equiv \delta(u_i)/\Delta\epsilon$.
Thus the allowed values of the elastic strain $\epsilon^{el}$ are $\epsilon_j = (-\ell_i,-\ell_i+1,...,\ell_i+k)\Delta \epsilon$.
The thermalized initial condition (Eq.\ \ref{eq:ic1}) becomes 
\small
\begin{equation}
w(u_i,\epsilon_j,t_0) = \bigg{\{ } 
\begin{array}{lll}
w_{eq}(u_i,\epsilon_j) \Delta u \Delta \epsilon & , & 0 \leq i \leq \alpha^2/\Delta u \\
& &  \rm{and}\ -\ell_i \leq j \leq \ell_i \\
& & \\
0 & , & i > \alpha^2/\Delta u \ \rm{or} \ |j| > \ell_i
\end{array}.
\label{eq:initialcond}
\end{equation}
\normalsize
Eq.\ \ref{eq:flow} is then integrated forward in time until $\epsilon$ reaches its final target value $\epsilon_{max}$.
A wide range of strain rates can be treated at fixed computational cost because our choice of timestep ($\Delta t = \Delta\epsilon/\dot\epsilon$) sets the required number of iterations of Eq.\ \ref{eq:flow}, $k_{max} = \epsilon_{max}/\Delta \epsilon$, to be independent of $\dot\epsilon$.
In general, the computational cost of deformation runs scales as $(\alpha^2/\Delta u)(\epsilon_{max}/\Delta\epsilon)^2 \delta(\alpha^2)$.
Here the numerical parameters $\Delta u = .01$ and $\Delta\epsilon = 10^{-5}$ were chosen to be small enough to achieve convergence of all results presented below.
The C++ code employed for all calculations presented herein is available online \cite{codelink}.

\begin{figure*}
\includegraphics[width=7in]{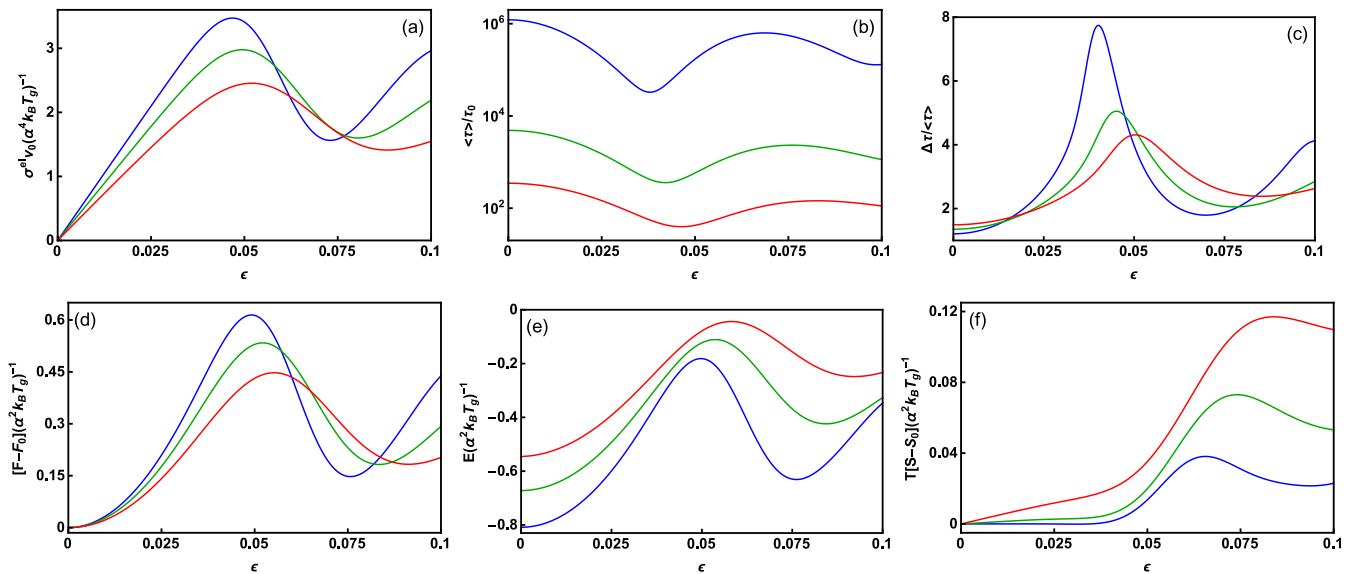}
\caption{Dependence of nonlinear mechanics and thermodynamics on $\epsilon$ and $T$ for thermalized SGR. Panel (a):\ Elastic stress-strain curves [$\sigma^{el}(\epsilon)$ (Eq.\ \ref{eq:stress})].  Panels (b-c):\ average zone relaxation time $\left< \tau \right>/\tau_0$ and its dispersion $\Delta \tau / \left< \tau \right>$. 
Panels (d-f):\ free energy $F(\epsilon)$, energy $E(\epsilon)$, and temperature$\times$entropy $TS(\epsilon)$  (Eqs.\ \ref{eq:E1}-\ref{eq:S1}). Blue, green, and red lines respectively indicate $T/T_g = 1/2$, $3/4$, and $39/40$.  Systems have $\alpha = \sqrt{8}$ and are deformed at constant strain rate $\dot{\epsilon} = \tau_0^{-1}$. All energies are scaled by the maximum zone activation energy $\alpha^2 k_B T_g$, and stresses are further scaled by $\alpha^2/v_0$. }
\label{fig:1}
\end{figure*}

\begin{table*}
\caption{Characteristic strains $\epsilon^y$ and $\epsilon^{pysm}$ and their associated elastic stresses $\sigma^y$ and $\sigma^{pysm}$ for thermalized (th) and nonthermalized (nth) $\alpha= \sqrt{8}$ systems.  $\epsilon^y$ and $\epsilon^{pysm}$ are respectively the strains at yield and at the postyield stress minimum (Figs.\ \ref{fig:1}a and \ref{fig:5}a).  $\sigma^y$ and $\sigma^{pysm}$ are scaled by $\alpha^4 k_B T_g/v_0$. Note that these results are for a high strain rate ($\dot\epsilon\tau_0 = 1$); their rate dependence is discussed in the Appendix.}
\begin{ruledtabular}
\begin{tabular}{lcccccccc}
$T/T_g$ & $\epsilon^y$ (th) & $\sigma^y$ (th) & $\epsilon^{pysm}$ (th) & $\sigma^{pysm}$ (th) & $\epsilon^y$ (nth) & $\sigma^y$ (nth) & $\epsilon^{pysm}$ (nth) & $\sigma^{pysm}$ (nth) \\
1/2 & 0.0468 & 3.472 & 0.0730 & 1.562 & 0.0550 & 4.521 & 0.0661 & 0.1636 \\
3/4 & 0.0494 & 2.977 & 0.0805 & 1.600 & 0.0576 & 4.122 & 0.0755 & 0.3124 \\
39/40 & 0.0519 & 2.454 & 0.0883 & 1.411 & 0.0602 & 3.721 & 0.0843 & 0.4483
\end{tabular}
\end{ruledtabular}
\label{tab:chareps}
\end{table*}

\section{Results for systems' mechanics, dynamics, and thermodynamics}

We now proceed to analyzing an example system's mechanics, dynamics and thermodynamics using the above formulae.
Figure \ref{fig:1} shows results for $\alpha = \sqrt{8}$ systems deformed at a high strain rate ($\dot{\epsilon} = \tau_0^{-1}$) to a maximum strain  $\epsilon_{max} = 0.1$.
With these parameters, deformation runs took no more than 8 hours on one CPU core.
Results are shown for three temperatures: $T/T_g = 1/2$, $3/4$, and $39/40$.
The first two are typical values of $T_{room}/T_g$ for metallic and polymeric glasses, while $T/T_g = 39/40$ is chosen to represent systems slightly below $T_g$.

Panel (a) shows results for the elastic component of stress, 
\begin{equation}
\sigma^{el}(\epsilon) =  \displaystyle\int_0^{\alpha^2} \displaystyle\int_{-\delta(u)}^{\delta(u) + \epsilon} \mathcal{K}_u \epsilon^{el} w(u,\epsilon^{el})  d\epsilon^{el} du.
\label{eq:stress}
\end{equation}
Following standard SGR-theoretic practice \cite{sollich98}, we focus on this elastic term.
Stress is simply an integral over contributions from different zones that are coupled only through the trap-model-style dynamics (Eqs.\ \ref{eq:trapflow}-\ref{eq:flow}) and thus interact only weakly.
The elastic response is temperature-dependent because systems at higher $T$ lie higher on their energy landscapes, i.e.\ the initial condition $w_{init}(u,\epsilon^{el}, T) = w_{eq}(u,\epsilon^{el},T)$ increasingly favors plastic zones with lower $u$ and hence lower $\mathcal{K}_u$ as $T$ increases.
Anelastic decrease of $\partial \sigma/\partial \epsilon$ sets in at lower strains and strengthens more with increasing $\epsilon$ at higher $T$.
Yield stresses $\sigma^y$ decrease with increasing $T$, while yield strains $\epsilon^y$ increase (Table \ref{tab:chareps}).
These temperature dependencies are relatively weak here because $\alpha^2 \gg 1$ and the applied strain rate is high ($\dot\epsilon\tau_0 = 1$).
Beyond yield, systems display dramatic strain softening that -- as in experiments \cite{schuh07,roth16} -- weakens with increasing $T$.
Within the present theory, the reason that strain softening weakens with increasing $T$ is as follows: at higher $T$, more zones yield at $\epsilon < \epsilon^y$, and hence fewer zones are in low-stress (small-$\epsilon^{el}$) states at $\epsilon = \epsilon^{pysm}$.
At still larger strains, a postyield stress minimum of the type observed in some metallic glasses \cite{johnson02,lu03,schuh07} is present at $\epsilon = \epsilon^{pysm}$.
This minimum occurs because yielding releases a large fraction of systems' elastic strain, which then builds up again as deformation continues.
See the Appendix for a discussion of how all of these effects vary with $\dot\epsilon\tau_0$.

Systems exhibit complex yielding \textit{dynamics}.
Panel (b) shows the average zone relaxation time (inverse yielding rate) $\left< \tau(\epsilon)\right>$.
$\left<\tau(\epsilon)\right>$ decreases rapidly with increasing strain as stress-activated plasticity becomes increasingly important, passes through a minimum at $\epsilon \simeq \epsilon^y$, increases again for $\epsilon > \epsilon^y$, and then decreases again for $\epsilon > \epsilon^{pysm}$.
All trends are consistent with experimental observations \cite{lee09,bending14,bending16} showing that relaxation in real glasses often speeds up by orders of magnitude near yielding, and can then slow down again upon strain softening.
Panel (c) shows the dynamical heterogeneity $\frac{\Delta \tau}{\left< \tau \right>} = \frac{\sqrt{  \left< \tau^2 \right> - \left< \tau \right>^2}}{\left< \tau \right>}$ of this relaxation.
Heterogeneity increases markedly with increasing strain for $\epsilon < \epsilon^y$, then decreases again for $\epsilon > \epsilon^y$.
The reason that heterogeneity increases is that plastic flow populates zones with an increasingly wide range of $u$ and $\epsilon^{el}$ as deformation proceeds, especially when many zones are yielding.
Note that such effects can be finely adjusted within the present model by varying the functional forms of $\mathcal{K}_u$ and $\delta(u)$.

Experiments typically show \cite{lee09,bending14,bending16} that heterogeneity \textit{decreases} during yielding and remains relatively low during plastic flow.
The different trends shown in panel (c) may arise because currently available SGR theories lack any ``facilitation'' mechanism.
Mechanical facilitation is the speedup of yielding that occurs in heterogeneous systems when zones have a broad distribution of stresses \cite{dequidt16}.
Zones that carry stresses much higher than the average value $\left< \sigma \right>$ yield faster because their environments cannot maintain local mechanical equilibrium (i.e.\ cannot force-balance such large stresses), and zones that carry very low stresses may similarly yield faster when $\left< \sigma \right>$ is large.
The net effect is homogenization of the yielding dynamics and of systems' relaxation in the postyield, plastic-flow regime  \cite{lee09,bending14,bending16,dequidt16}.
It would be interesting in future work to add mean-field facilitation (or a comparable stress-diffusion mechanism \cite{hebraud98}) to SGR theory.

In Eqs.\ \ref{eq:E1}-\ref{eq:tauinv}, the zone populations $w(u,\epsilon^{el})$ are strain-history-dependent.
For systems that have undergone plastic deformation, $w(u,\epsilon^{el}) \neq w_{eq}(u,\epsilon^{el})$, and $E(\epsilon)$, $S(\epsilon)$, and $F(\epsilon)$ are not thermodynamic state functions, but instead are inherently nonequilibrium quantities.
It is therefore worthwhile to examine their evolution during deformation. 
Panels (d-f) of Fig.\ \ref{fig:1} show $F(\epsilon)$, $E(\epsilon)$, and the entropic component of free energy $TS(\epsilon)$. 
To facilitate comparison of systems at different $T$, $\Delta F = F(\epsilon) - F(0)$ and $T\Delta S = T[S(\epsilon) - S(0)]$ are shown rather than the bare values, and all quantities are scaled by the characteristic energy $\mathcal{U}_{max} = \alpha^2 k_B T_g$.
As expected, results for $F$ and $E$ are quadratic in strain in the elastic regime, reach maxima near $\epsilon^y$, then decrease during strain softening as massive zone yielding releases stored elastic strain energy.
The dominant contribution to $\partial F/\partial\epsilon$ is energetic as long as $\dot\epsilon \left< \tau \right>$ is large.
Thermalized flow (Eqs.\ \ref{eq:trapflow}-\ref{eq:flow}) creates new zones with probabilities proportional to their equilibrium statistical weights $w_{eq}(u,\epsilon^{el},T)$, and therefore acts to push systems back towards their initial states, producing the negative $\partial S/\partial \epsilon$ for $\epsilon > \epsilon^y$.
Note that similar calculations of $S(\epsilon)$ cannot be straightforwardly performed in discrete-zone implementations of SGR \cite{warren08,fielding14,merabia16,radhakrishnan17} because they do not explicitly determine $p(u,\epsilon^{el})$ as we do here.

\begin{figure}[htbp]
\includegraphics[width=3in]{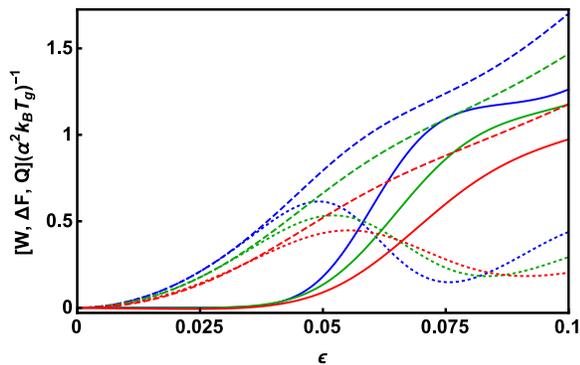}
\caption{Energy dissipation in the same systems depicted in Fig.\ \ref{fig:1}.  Dotted, dashed, and solid lines respectively show $\Delta F(\epsilon)$, $W(\epsilon)$, and $Q(\epsilon)$.  Note that the stress $\sigma(\epsilon)$ appearing in the definition of $W(\epsilon) = \int_0^\epsilon \sigma(\epsilon')d\epsilon'$ is $\sigma(\epsilon) = \sigma^{el}(\epsilon) + \sigma^{entr}(\epsilon)$, where the entropic term $\sigma^{entr}(\epsilon) = -T(\partial S/\partial \epsilon)$.}
\label{fig:2}
\end{figure}

\begin{figure*}
\includegraphics[width=7in]{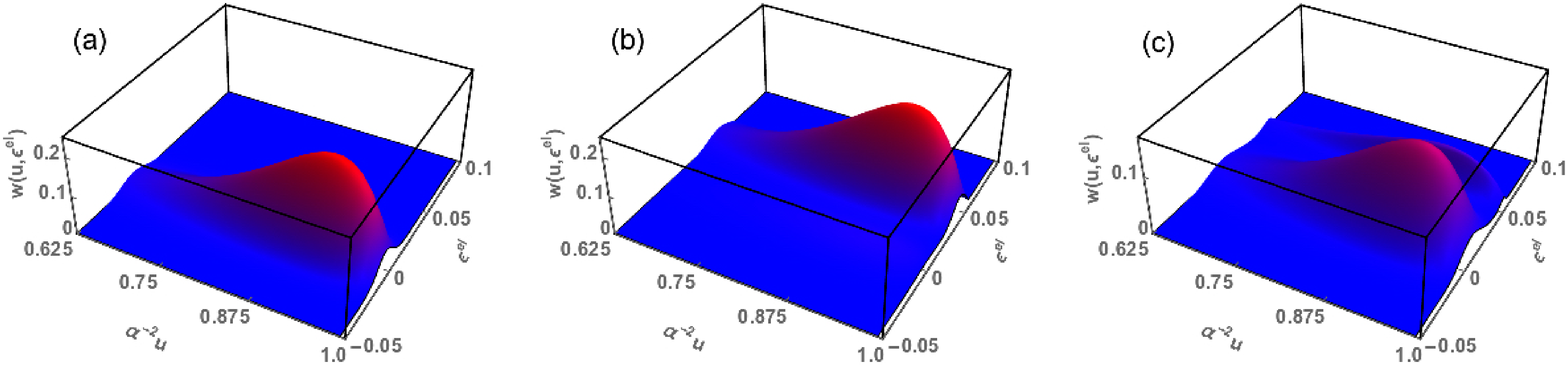}
\caption{Strain-history-dependent position of the thermalized $T/T_g = 3/4$ system on its energy landscape. Panel (a): $w(u,\epsilon^{el})$ for unstrained systems ($\epsilon = 0$).  Panel (b): $w(u,\epsilon^{el})$ at the yield strain ($\epsilon = \epsilon^y = 0.0468$).  Panel (c): $w(u,\epsilon^{el})$ at the postyield stress minimum ($\epsilon =  \epsilon^{pysm} = 0.0730$).  Because this $\alpha=\sqrt{8}$ system remains low on its energy landscape, $w(u,\epsilon^{el})$ is shown only for $3/4 \leq \alpha^2 u \leq 1$; values for $\alpha^2 u < 3/4$ are finite but remain small.}
\label{fig:3}
\end{figure*}

Since the mechanical work $W(\epsilon) = \int_0^\epsilon \sigma(\epsilon')d\epsilon'$ satisfies the first law of thermodynamics, its dissipated component is simply $Q(\epsilon) \equiv W(\epsilon) - \Delta F(\epsilon)$, where $\Delta F = F(\epsilon)-F(0)$.
Figure \ref{fig:2} shows that as expected, $Q(\epsilon)$ is small in the elastic regime, but grows rapidly at larger strains.
Rapid growth of $Q(\epsilon)$ begins in the anelastic regime, as ``softer'' \cite{riggleman10b,manning11,ding14,schoenholz14} zones [zones with lower activation energies and yield strains] begin yielding.
For $\epsilon_y \lesssim \epsilon \lesssim \epsilon^{pysm}$, most mechanical work is dissipated.
One might naively suppose the large $Q(\epsilon)$ to be at logical odds with the relatively small increase in $S(\epsilon)$.
However, this is not so, because the thermalized plastic flow rule (Eqs.\ \ref{eq:trapflow}-\ref{eq:flow}) causes flow to be mostly into less-strained zones of similar $u$, i.e.\ to push $w(u,\epsilon^{el})$ back towards its initial state.
We will show below (Sec.\ \ref{sec:nonthermal}) that nonthermalized flow produces strikingly different behavior.

In real systems, the high levels of energy dissipation depicted in Fig.\ \ref{fig:2} often produce temperature increases that in turn lead to enhanced strain softening \cite{arruda95,schuh07}. 
Here, for simplicity, we ignore such effects and assume that the coupling of systems to their environmental thermal reservoirs maintains constant $T$.
This is a potentially inaccurate approximation, and should be corrected in future work when necessary.
Any such corrections, however, will presumably \cite{fuereder13} require system-specific treatments that are beyond the scope of this study.
Here we have shown $Q(\epsilon)$ in Fig.\ \ref{fig:2} mainly to motivate what follows.

Our method's prediction of the strain-history-dependent zone statistical weights $w(u,\epsilon^{el};\epsilon)$ allows easy visualization of how systems' positions on their energy landscape evolve during deformation \cite{footvias}.
Elastic (plastic) deformation can then be identified with affine (nonaffine) evolution of $w(u,\epsilon^{el};\epsilon)$.
Concurrently, since the thermalized plastic flow rule (Eqs. \ref{eq:trapflow}-\ref{eq:flow}) is consistent with the standard thermodynamic identification of dissipated work as \textit{heat} that changes microstate populations, 
$Q(\epsilon)$ can be directly related to systems' flow over their energy landscapes and hence to the  fundamental character of their plastic flow.
Figure \ref{fig:3} shows $w(u,\epsilon^{el};\epsilon)$ for the $T/T_g= 1/2$ system at the three representative strains (Table \ref{tab:chareps}) $\epsilon = 0$, $\epsilon = \epsilon^y$, and $\epsilon = \epsilon^{pysm}$.
Panel (a) shows the initial $\epsilon = 0$ distribution, which illustrates how thermalization of strain degrees of freedom influences systems' initial positions on their energy landscapes.
Notably, many zones have initial elastic strains $\epsilon^{el}_{init}$ that are not negligible compared to their yield strains $\epsilon^y_u$.
Panel (b) shows that the majority of the zones present at $\epsilon = 0$ deform affinely (i.e.\ do not yield) throughout the strain range $0 \leq \epsilon \leq \epsilon^y$.
Those that do yield by $\epsilon = \epsilon^y$ are primarily those with positive $\epsilon^{el}_{init}$.
In contrast, by $\epsilon = \epsilon^{pysm}$, most zones have yielded and been replaced by new zones with smaller $\epsilon^{el}$.
This can be seen in panel (c):\ the primary maximum of $w(u,\epsilon^{el}; \epsilon^{pysm})$ is at $\epsilon^{el} < \epsilon^y$.
However, the secondary maximum at $\epsilon^{el} > \epsilon^y$ shows that some of the zones present in the initial undeformed state remain intact at $\epsilon = \epsilon^{pysm}$.
As expected, most of these had negative $\epsilon^{el}_{init}$.
These results are closely connected to the breadth of the stress maxima shown in Fig.\ \ref{fig:1}(a); all reflect the fact that yielding is a gradual process.

Comparing panels (a-c) of Fig.\ \ref{fig:3} helps us understand how mechanical work gets dissipated during strain softening.
Elastic strain energy gets released as plastic zones yield.
Thermalized plastic flow takes the system back towards its initial position on its energy landscape.
Indeed, its position at  $\epsilon = \epsilon^{pysm}$ is closer to its initial $\epsilon = 0$ position than to its position at  $\epsilon = \epsilon^y$, consistent with the $F(\epsilon^{pysm}) - F(0) < F(\epsilon^y) - F(\epsilon^{pysm}) < F(\epsilon^y) - F(0)$ result [as well as similar trends in $E(\epsilon)$ and $S(\epsilon)$] shown in Figure \ref{fig:1}(d-f). 
It will be interesting in future work to repeat this exercise for different $\alpha$ as well as different initial conditions, e.g.\ nonequilibrium $w_{init}(u,\epsilon^{el}) \neq w_{eq}(u,\epsilon^{el})$ that more accurately reflect typical glasses; cf.\ Sec.\ \ref{sec:conclude}.

\begin{figure}
\includegraphics[width=3in]{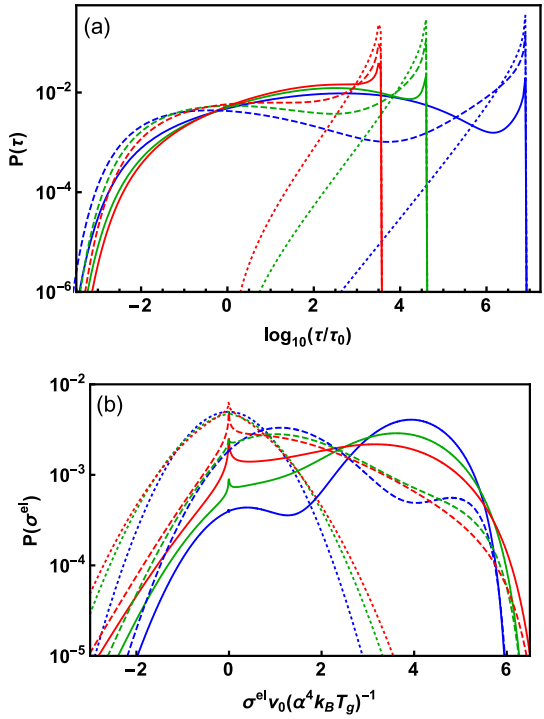}
\caption{Strain-history-dependent probability distributions of zones' relaxation times [$P(\tau; \epsilon)$ from Eq.\ \ref{eq:poftau}:\ panel (a)] and elastic stresses [$P(\sigma^{el}; \epsilon)$ from Eq.\ \ref{eq:pofsigmael}:\ panel (b)].  Systems are the same and line colors indicate temperatures as in Figs.\ \ref{fig:1}-\ref{fig:2}.  Dotted, solid, and dashed curves respectively indicate data for $\epsilon = 0$, $\epsilon = \epsilon^y$, and $\epsilon^{pysm}$ (Table \ref{tab:chareps}).}
\label{fig:4}
\end{figure}

The $w(u,\epsilon^{el}; \epsilon)$ distributions calculated by integrating Eq.\ \ref{eq:flow} enable prediction of many other physical properties, such as strain-history-dependent probability distributions of zone relaxation times
\begin{equation}
P(\tilde\tau; \epsilon) = \displaystyle\int_0^{\alpha^2} \displaystyle\int_{-\delta(u)}^{\delta(u)+\epsilon} w(u,\epsilon^{el}) \delta[\tilde\tau - \tau(u,\epsilon^{el}, T)]d\epsilon^{el} du,
\label{eq:poftau}
\end{equation}
elastic stresses
\begin{equation}
P(\tilde{\sigma}^{el}; \epsilon) = \displaystyle\int_0^{\alpha^2} \displaystyle\int_{-\delta(u)}^{\delta(u)+\epsilon} w(u,\epsilon^{el}) \delta[\tilde{\sigma}^{el} - \mathcal{K}_u\epsilon^{el}]d\epsilon^{el} du,
\label{eq:pofsigmael}
\end{equation}
and any other relevant thermodynamic or mechanical quantity.
Figure \ref{fig:4} shows results for $P(\tau; \epsilon)$ and $P(\sigma^{el}; \epsilon)$ for the three characteristic strains $\epsilon=0$, $\epsilon = \epsilon^y$, and $\epsilon = \epsilon^{pysm}$ (Table \ref{tab:chareps}).
Zone yielding rates span many orders of magnitude because of their wide ranges of $u$ and $\epsilon^{el}$.
Temperature and stress/strain affect the shapes of the $P(\tau)$ distributions in nontrivial ways.
For example, stress-activated relaxation and plastic deformation do not merely shift $\left<\tau\right>$ or transform the $P(\tau)$ distributions in any ``affine'' manner; instead, they change their shape, as can be seen by comparing the distributions for $\epsilon=\epsilon^y$ and   $\epsilon=\epsilon^{pysm}$.
Even more complex physics can be seen in the $P(\sigma)$ distributions.
At $\epsilon = 0$, they are simply Gaussian distributions reflecting the thermalized initial condition (Eq.\ \ref{eq:ic1}).
In contrast, the split peaks of the $T/T_g = 1/2$ distributions for $\epsilon = \epsilon^{y}$ and $\epsilon = \epsilon^{pysm}$ reflect the emerging coexistence of yielded and unyielded plastic zones depicted in Fig.\ \ref{fig:3}.

Distributions like $P(\tau; \epsilon)$ and $P(\sigma; \epsilon)$ contain much information that cannot be gleaned from their mean values ($\left< \tau(\epsilon) \right>$ or $\sigma^{el}(\epsilon)$).
For example, the tails of the distributions may dominate certain physical phenomena such as aging during deformation \cite{warren08,roth16}, and may therefore be important for understanding the mechanics of heterogeneous systems (i.e.\ glasses) in more detail. 
Clearly, discrete-zone implementations of SGR (e.g.\ \cite{warren08,fielding14,merabia16,radhakrishnan17}) cannot easily provide distributions spanning many orders of magnitude in probability as we have done here.

\begin{figure*}
\includegraphics[width=6.9in]{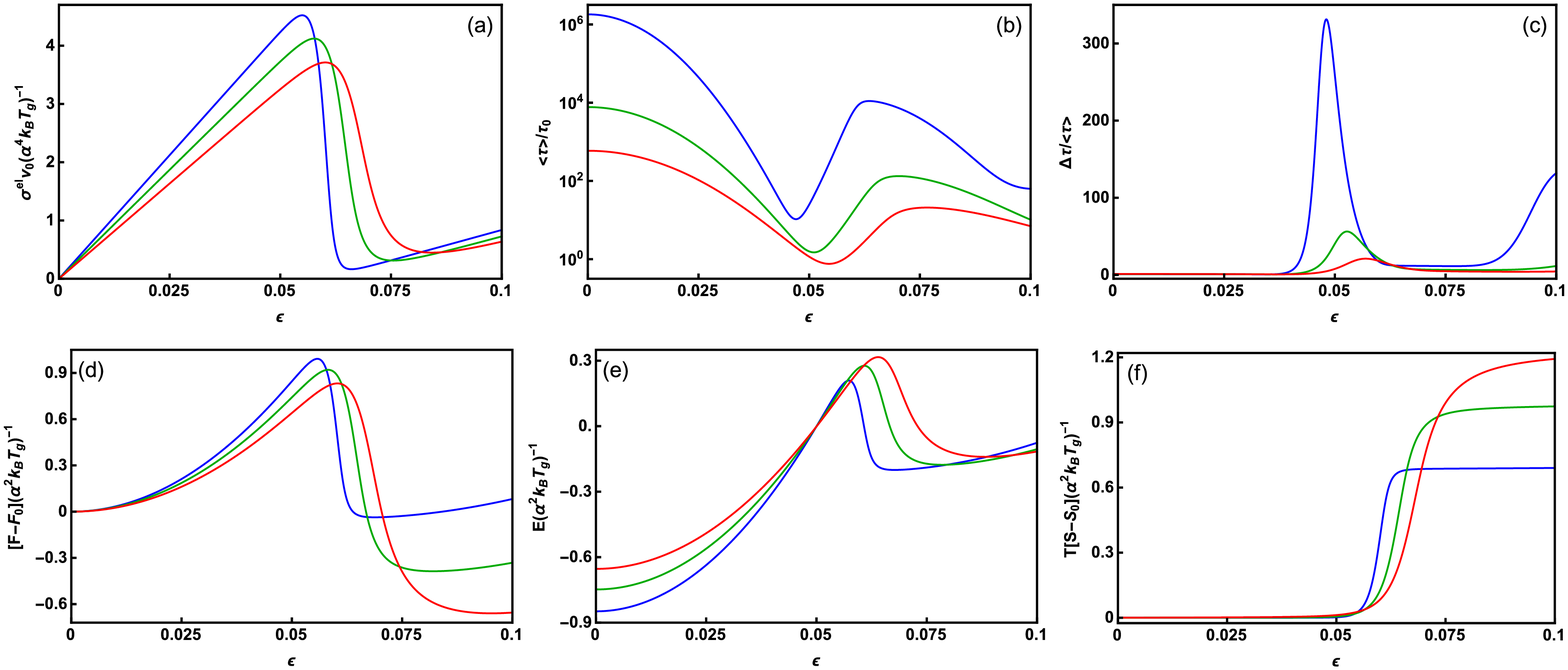}
\caption{Dependence of nonlinear mechanics and thermodynamics on $\epsilon$ and $T$ for  nonthermalized SGR. Panel (a):\ Elastic stress-strain curves [$\sigma^{el}(\epsilon)$ (Eq.\ \ref{eq:stress})].  Panels (b-c):\ average zone relaxation time $\left< \tau \right>/\tau_0$ and its dispersion $\Delta \tau / \left< \tau \right>$. 
Panels (d-f):\ free energy $F(\epsilon)$ , energy $E(\epsilon)$, and temperature$\times$entropy $TS(\epsilon)$ (Eqs.\ \ref{eq:E1}-\ref{eq:S1}). Blue, green, and red lines respectively indicate $T/T_g = 1/2$, $3/4$, and $39/40$.  Systems have $\alpha = \sqrt{20}$ and are deformed at constant strain rate $\dot{\epsilon} = \tau_0^{-1}$. All energies are scaled by the maximum zone activation energy $\alpha^2 k_B T_g$, and stresses are further scaled by $\alpha^2/v_0$. Note that replacing the $\delta(\epsilon^{el})$ proportionality in Eq.\ \ref{eq:stdflow} with a $\theta(u,\epsilon^{el})$ proportionality eliminates the post-softening ($\epsilon > \epsilon^{pysm}$) stress increases shown in panel (a); instead, systems exhibit perfect-plastic flow at a constant stress $\sigma_{flow}(T)$, which in turn affects other measures of response like those shown in panels (b-f).}
\label{fig:5}
\end{figure*}

\section{Comparison to nonthermalized SGR}
\label{sec:nonthermal}

To illustrate the significance of thermalization, we contrast some of the above results to those obtained from a more traditional version of SGR theory where (as in the original formulation \cite{sollich97,sollich98})
strain degrees of freedom are not thermalized and newly created zones have zero strain. 
As in Refs.\ \cite{sollich97,sollich98,fielding00,warren08,fielding14,merabia16,radhakrishnan17}, we assume all zones are initially unstrained.
Then the equilibrated initial condition then becomes $w(u,\epsilon^{el},T) = \rho(u) p_{eq}(u,0,T)\delta(\epsilon^{el})$, the allowed values of $j$ at time $t_k$ are ($0,1,...,k$), and the allowed values of the elastic strain $\epsilon^{el}$ are $\epsilon_j = (0,1,...,k)\Delta \epsilon$.
The evolution equation for zone populations (i.e.\ the traditional-SGR counterpart to Eq.\ \ref{eq:flow}) is  
\begin{equation}
\begin{array}{c}
z(u_i, \epsilon_j; t_k) = z(u_i, \epsilon_{j-1}; t_{k-1}) \left[1 -\displaystyle\frac{\Delta t}{\tau(u_i,\epsilon_{j-1},T)}\right] \\
\\
+\ \rho(u)  \delta( \epsilon_j ) \left< \tau^{-1} (\epsilon_{j-1}) \right>  \Delta t.
\end{array}
\label{eq:stdflow}
\end{equation}
On the right hand side of Eq.\ \ref{eq:stdflow}, the second term indicates standard SGR-style creation \cite{sollich97,sollich98} of new unstrained zones; the $p_{eq}(u,\epsilon^{el},T)$ factors present in Eq.\ \ref{eq:flow} are absent here because traditional SGR is a nearly athermal theory.
The other equations (\ref{eq:E1}, \ref{eq:S1}, \ref{eq:tauinv}, \ref{eq:stress}) for thermodynamics, dynamics, and mechanics we have used above remain the same, but the different assumptions made by traditional SGR impose $\delta(u) = 0$  [in contrast to the $\delta(u) = 1/\sqrt{k(u)}$ condition derived in Sec.\ \ref{sec:theory}].

These different theoretical assumptions produce a considerably different physical response.
Figure \ref{fig:5} shows the traditional-SGR counterparts to the results shown in Fig.\ \ref{fig:1}.
Panel (a) shows that nonthermalized systems' yielding behavior differs in several ways from their thermalized counterparts:\ (i) their yield strains $\epsilon^y$ and yield stresses $\sigma^y$ are larger; (ii) their anelastic regime is narrower; and (iii) their postyield strain softening is much sharper, occurring over a narrower strain window and ending at a lower stress minimum; see Table \ref{tab:chareps}.
The extremely low values of $\sigma^{pysm}$ arise because the combination of the traditional-SGR initial condition ($\epsilon^{el}_{init }= 0$ for all zones) with our chosen $k(u)$ [that produces $\epsilon^y_u = .05$ for all $u$] means that all zones yield nearly simultaneously.
As a consequence of effect (i), average relaxation times [panel (b)] drop more in nonthermalized systems than in their thermalized counterparts.
Differences in the dynamical heterogeneity of yielding [panel (c)] arise because nonthermalized plastic flow (Eq.\ \ref{eq:stdflow}) populates zones with a wider range of $u$ -- and hence a wider range of $\tau$ -- than its thermalized counterpart (Eq.\ \ref{eq:flow}.

As shown in panels (d-f), the nonequilibrium thermodynamics of deformation are also quite different in traditional SGR. 
 This is a consequence of both the different initial states of systems and the different plastic flow rules.
 In thermalized systems, the strain energy in zones with negative $\epsilon^{el}_{init}$ decreases with increasing $\epsilon$ for $\epsilon < |\epsilon^{el}_{init}|$.
No such zones are present in nonthermalized systems.
This causes nonthermalized systems to be driven much further up their energy landscapes prior to yielding than they are in their thermalized counterparts, so much so that for high strain rates an unstable ($E(\epsilon) > 0$) flow regime appears for $\epsilon \gtrsim .05$ \cite{footEgt0}.
Moreover, nonthermalized plastic flow populates the upper regions of systems' energy landscapes much more than its thermalized counterpart.
Specifically, it produces both the much higher $E(\epsilon)$ for $\epsilon \gtrsim \epsilon^y$ shown in panel (e) and the massive entropy increase at $\epsilon \simeq \epsilon^y$ shown in panel (f).
The magnitude of $\Delta S(\epsilon)$ is so large partially because zones' elastic strains are nearly $\delta$-function distributed [i.e.\ $w(u,\epsilon^{el}; \epsilon) \propto \delta(\epsilon^{el} - \epsilon)$] for $\epsilon \ll \epsilon^y$, but become broadly distributed for $\epsilon \gtrsim \epsilon^y$.
Replacing the initial condition [$w(u,\epsilon^{el},T) = w_{eq}(u,0,T)\delta(\epsilon^{el})$] with a strain-thermalized initial condition (Eq.\ \ref{eq:ic1}) or replacing the flow factor $f(u,\epsilon^{el}) = \rho(u)\delta(\epsilon^{el})$ in Eq.\ \ref{eq:stdflow} with a nonthermalized version of Eq.\ \ref{eq:fofx} reduces $\Delta S$ considerably.
However, $\Delta S$ remains considerably larger than it is for thermalized flow because (as noted above) nonthermalized flow populates 
lower-$u$ zones to a much greater degree.
This combination of sharp increases in $E$, $S$ and mobility (i.e.\ $\left< \tau^{-1} \right>$) in nonthermalized systems is consistent with the old idea \cite{robertson66} that yielding effectively ``melts'' glasses.
Also consistent with this idea is the fact that beyond yield, $E(\epsilon)$ depends only weakly on $T$.
That these behaviors are present for nonthermalized but not for thermalized plasticity is of considerable interest.

\section{Discussion and Conclusions}
\label{sec:conclude}

Many modern theories of plasticity, including the SGR and STZ theories, employ effective temperatures $T_{eff}$ to reflect the fact that the slowly relaxing configurational degrees of freedom (i.e.\ plastic zones) in deforming systems tend to fall out of equilibrium with their fast kinetic/vibrational degrees of freedom and with their environmental thermal reservoir.
They reason that $T_{eff}$, which is thermodynamically conjugate \cite{cugliandolo97,berthier00} to the configurational entropy associated with the plastic zones, in general differs from the reservoir temperature $T$.
Recent thermodynamics-focused work \cite{bouchbinder09a,bouchbinder09b,langer12,sollich12,bouchbinder13,fuereder13,kamrin14} has shown how to rigorously account for energy and entropy transfer between these slow and fast degrees of freedom, and hence to predict the evolution of $T_{eff}$ during deformation.

The formulation of SGR theory developed in Sec.\ \ref{sec:theory} allowed us to straightforwardly take the alternative approach of directly calculating systems' strain-history-dependent positions on their energy landscapes, i.e.\ $w(u, \epsilon^{el}; \epsilon)$ distributions like those illustrated in Fig.\ \ref{fig:3}.
Since it allows for direct calculation of $w(u,\epsilon^{el}; \epsilon)$, the present theory has no need for a $T_{eff}$-like quantity.
Similarly, since in contrast to standard SGR (where $x$ reflects the degree to which flow is thermalized by mechanical ``kicks'' \cite{sollich97,sollich98,fielding00} from surrounding zones) the present theory assumes that these kicks are themselves thermalized by systems' fast degrees of freedom and therefore that their magnitude is set by the reservoir temperature $T$, it has no need for any $x$-like quantity.
While its ``$T$-only'' approach probably restricts its applicability to the most thermal amorphous materials, e.g.\ metallic, small-molecule, and polymeric glasses, such materials are both commonplace and technologically important.

SGR theory assumes that the boundaries between basins on systems' potential energy landscapes lie at $\mathcal{U}=0$, independent of $u$.  
This assumption can be used to justify both the traditional SGR flow law (Eq.\ \ref{eq:stdflow}) and the thermalized version (Eq.\ \ref{eq:flow})  discussed herein.
Eq.\ \ref{eq:stdflow} is obtained by assuming that when zones yield, they are replaced by new zones that randomly (athermally) populate  basins on the system's energy landscape.
In contrast, Eq.\ \ref{eq:flow} assumes that new zone selection is fully thermalized, i.e.\ new zones populate basins with probability proportional (Eq.\ \ref{eq:fofx}) to their equilibrium occupation probability $p_{eq}(u,\epsilon^{el},T)$.
Which flow law is more realistic for a given system will depend on the degree to which the system is thermal -- i.e., upon $\dot\epsilon \tau_0$, the ratio of $k_B T$ to the system's mechanically relevant energy scales, as well as other factors \cite{langer12}, in some presumably complicated fashion -- and the behavior of real systems lies, in all likelihood, somewhere between these two limiting cases.
Here our purpose was not to determine where any specific system lies along the athermal--thermal continuum, but simply to illustrate in a pedagogical way various consequences of the differences between the physics assumptions used to derive Eqs.\ \ref{eq:flow} and \ref{eq:stdflow}.

Consistent with this purpose, we made two further simplifying approximations.
First, following SGR-theoretic convention \cite{sollich97,sollich98,fielding00,sollich12,bouchbinder13,fuereder13,warren08,fielding14,merabia16,radhakrishnan17}, we treated plastic zones as internally structureless.
Since plastic zones in real systems are composed of the systems' constituent particles and their internal entropy consequently tends to decrease with increasing strain, the present theory may need to be modified to incorporate a strain-dependent density-of-states function [$\rho^*(u,\epsilon^{el})$] to optimally model real materials.
Such modifications will be highly system-specific -- for example, the $\epsilon^{el}$-dependence of $\rho^*(u,\epsilon^{el})$ will be different for polymeric vs.\ metallic glasses \cite{schuh07,roth16} -- and are therefore beyond the scope of this initial study.
Second, the equilibrated initial condition employed here [$w(u,\epsilon^{el}; 0) = w_{eq}(u,\epsilon^{el})$]  is obviously an idealization that is not physically representative of most real glasses.
We chose it to set up an easily understood demonstration of the present formulation's potential for analyzing systems' deformation thermodynamics, and in particular their plastic flow over their energy landscapes.
However, we emphasize that all methods described herein can be employed with arbitrary initial conditions.
For example, to model ``young'' glasses, one can set $w_{init}(u, \epsilon^{el}) = \rho(u)p_{eq}(u,\epsilon^{el},T_{eff})$ with a $T_{eff} > T$ that slowly approaches $T$ during the aging process \cite{fielding00,bouchbinder09a,bouchbinder09b,sollich12,langer12,bouchbinder13,fuereder13,kamrin14,cugliandolo97}.

Ronald G.\ Larson, Gregory B.\ McKenna, Ken Kamrin, Samy Ferabia, Suzanne M.\ Fielding and Mark D.\ Ediger provided helpful discussions.
This material is based upon work supported by the National Science Foundation under Grant No.\ DMR-1555242.

\section{Appendix: Rate-dependent yielding and strain softening}

\begin{figure}[htbp]
\includegraphics[width=3in]{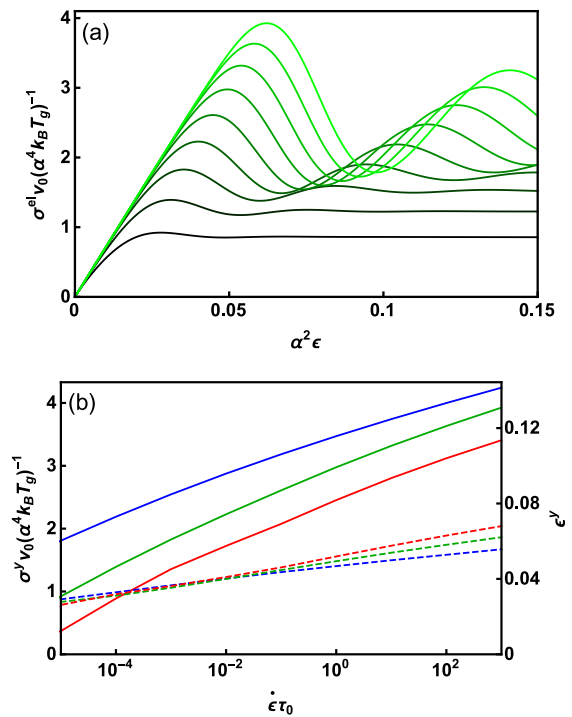}
\caption{Rate-dependent mechanical response for thermalized flow.  Panel (a): Stress-strain curves for $T/T_g = 3/4$, for strain rates $10^{-5} \leq \dot\epsilon\tau_0 \leq 10^3$; brighter green indicates higher rates.  Panel (b): Scaled yield stresses $\sigma^y(\dot\epsilon)$ [solid curves] and yield strains $\epsilon^y(\dot\epsilon)$ [dashed curves].  Systems are the same and line colors indicate temperatures as in Figs.\ \ref{fig:1}-\ref{fig:2}.}
\label{fig:6}
\end{figure}

One of the most common applications of plasticity theory has been prediction of the temperature and strain-rate dependences of systems' yield and flow stresses.
Here we report and discuss the $\dot\epsilon$-dependence of the various $T$-dependent quantities presented in Table \ref{tab:chareps}.
Figure \ref{fig:6} shows stress-strain curves for thermalized plastic flow at $T/T_g = 3/4$ over a wide range of strain rates $10^{-5} \leq \dot\epsilon\tau_0 \leq 10^3$.
All results are qualitatively consistent with trends observed in experimental studies of rate-dependent mechanical response in bulk metallic glasses \cite{johnson02,lu03,schuh07}.
Anelastic response sets in at lower strain for lower $\dot\epsilon$ because zones have more time to yield (via thermal activation) over any given strain interval.
Yield stresses and strains increase steadily with increasing strain rate.
Panel (b) shows that these increases are approximately logarithmic in $\dot\epsilon$, as is expected for thermally activated yielding \cite{ree55}.

Two features of the data shown in Fig.\ \ref{fig:6} are particularly noteworthy.
First, the temperature and rate dependencies of $\epsilon^y$ are strongly coupled; $\epsilon^y$ decreases with increasing $T$ at low $\dot\epsilon$, but increases with increasing $T$ at high $\dot\epsilon$.
This behavior reflects the fact that low-$\dot\epsilon$ yielding is primarily thermally-activated (i.e.\ driven by slow thermal activation over relatively large energy barriers), whereas high-$\dot\epsilon$ yielding is primarily stress-activated (i.e.\ the higher strain energies associated with the larger $\epsilon^y$ lower energy barriers and speed yielding).
Note that both positive and negative $\partial\epsilon^y/\partial T$ are observed in real glasses \cite{roth16,schuh07}.
Second, the strain rate dependence of $\sigma^{pysm}$ is far weaker than that of $\sigma^y$.
Stress overshoots (i.e.\ finite $|\sigma^y - \sigma^{pysm}|$) that increase with $\dot\epsilon$ are observed in a wide range of glassy materials, including most polymeric and metallic glasses \cite{roth16,schuh07}.
Analysis of the thermodynamics [$E(\epsilon)$, $S(\epsilon)$ and $F(\epsilon)$] shows that while systems deformed at higher strain rates are driven further up their energy landscapes, they return to similar positions on their energy landscapes [i.e.\ have comparable $w(x,\epsilon^{el}; \epsilon)$] at $\epsilon = \epsilon^{pysm}$ even though the values of $\epsilon^{pysm}$ are significantly different.
Results like this illustrate the utility of plasticity theories with properly thermalized strain degrees of freedom.

%\bibliography{plast}

\begin{thebibliography}{47}%
\makeatletter
\providecommand \@ifxundefined [1]{%
 \@ifx{#1\undefined}
}%
\providecommand \@ifnum [1]{%
 \ifnum #1\expandafter \@firstoftwo
 \else \expandafter \@secondoftwo
 \fi
}%
\providecommand \@ifx [1]{%
 \ifx #1\expandafter \@firstoftwo
 \else \expandafter \@secondoftwo
 \fi
}%
\providecommand \natexlab [1]{#1}%
\providecommand \enquote  [1]{``#1''}%
\providecommand \bibnamefont  [1]{#1}%
\providecommand \bibfnamefont [1]{#1}%
\providecommand \citenamefont [1]{#1}%
\providecommand \href@noop [0]{\@secondoftwo}%
\providecommand \href [0]{\begingroup \@sanitize@url \@href}%
\providecommand \@href[1]{\@@startlink{#1}\@@href}%
\providecommand \@@href[1]{\endgroup#1\@@endlink}%
\providecommand \@sanitize@url [0]{\catcode `\\12\catcode `\$12\catcode
  `\&12\catcode `\#12\catcode `\^12\catcode `\_12\catcode `\%12\relax}%
\providecommand \@@startlink[1]{}%
\providecommand \@@endlink[0]{}%
\providecommand \url  [0]{\begingroup\@sanitize@url \@url }%
\providecommand \@url [1]{\endgroup\@href {#1}{\urlprefix }}%
\providecommand \urlprefix  [0]{URL }%
\providecommand \Eprint [0]{\href }%
\providecommand \doibase [0]{http://dx.doi.org/}%
\providecommand \selectlanguage [0]{\@gobble}%
\providecommand \bibinfo  [0]{\@secondoftwo}%
\providecommand \bibfield  [0]{\@secondoftwo}%
\providecommand \translation [1]{[#1]}%
\providecommand \BibitemOpen [0]{}%
\providecommand \bibitemStop [0]{}%
\providecommand \bibitemNoStop [0]{.\EOS\space}%
\providecommand \EOS [0]{\spacefactor3000\relax}%
\providecommand \BibitemShut  [1]{\csname bibitem#1\endcsname}%
\let\auto@bib@innerbib\@empty
%</preamble>
\bibitem [{\citenamefont {Ree}\ and\ \citenamefont {Eyring}(1955)}]{ree55}%
  \BibitemOpen
  \bibfield  {author} {\bibinfo {author} {\bibfnamefont {T.}~\bibnamefont
  {Ree}}\ and\ \bibinfo {author} {\bibfnamefont {H.}~\bibnamefont {Eyring}},\
  }\bibfield  {title} {\enquote {\bibinfo {title} {Theory of non-newtonian
  flow. i. solid plastic system},}\ }\href@noop {} {\bibfield  {journal}
  {\bibinfo  {journal} {J. Appl. Phys.}\ }\textbf {\bibinfo {volume} {26}},\
  \bibinfo {pages} {793} (\bibinfo {year} {1955})}\BibitemShut {NoStop}%
\bibitem [{\citenamefont {Stillinger}(1995)}]{stillinger95}%
  \BibitemOpen
  \bibfield  {author} {\bibinfo {author} {\bibfnamefont {F.~H.}\ \bibnamefont
  {Stillinger}},\ }\bibfield  {title} {\enquote {\bibinfo {title} {A
  topographic view of supercooled liquids and glass-formation},}\ }\href@noop
  {} {\bibfield  {journal} {\bibinfo  {journal} {Science}\ }\textbf {\bibinfo
  {volume} {267}},\ \bibinfo {pages} {1935} (\bibinfo {year}
  {1995})}\BibitemShut {NoStop}%
\bibitem [{\citenamefont {Debenedetti}\ and\ \citenamefont
  {Stillinger}(2001)}]{debenedetti01}%
  \BibitemOpen
  \bibfield  {author} {\bibinfo {author} {\bibfnamefont {P.~G.}\ \bibnamefont
  {Debenedetti}}\ and\ \bibinfo {author} {\bibfnamefont {F.~H.}\ \bibnamefont
  {Stillinger}},\ }\bibfield  {title} {\enquote {\bibinfo {title} {Supercooled
  liquids and the glass transition},}\ }\href@noop {} {\bibfield  {journal}
  {\bibinfo  {journal} {Nature}\ }\textbf {\bibinfo {volume} {410}},\ \bibinfo
  {pages} {259} (\bibinfo {year} {2001})}\BibitemShut {NoStop}%
\bibitem [{\citenamefont {Tsamados}\ \emph {et~al.}(2009)\citenamefont
  {Tsamados}, \citenamefont {Tanguy}, \citenamefont {Goldenberg},\ and\
  \citenamefont {Barrat}}]{tsamados09}%
  \BibitemOpen
  \bibfield  {author} {\bibinfo {author} {\bibfnamefont {M.}~\bibnamefont
  {Tsamados}}, \bibinfo {author} {\bibfnamefont {A.}~\bibnamefont {Tanguy}},
  \bibinfo {author} {\bibfnamefont {C.}~\bibnamefont {Goldenberg}}, \ and\
  \bibinfo {author} {\bibfnamefont {J.-L.}\ \bibnamefont {Barrat}},\ }\bibfield
   {title} {\enquote {\bibinfo {title} {Local elasticity map and plasticity in
  a model lennard-jones glass},}\ }\href@noop {} {\bibfield  {journal}
  {\bibinfo  {journal} {Phys, Rev. E}\ }\textbf {\bibinfo {volume} {80}},\
  \bibinfo {pages} {026112} (\bibinfo {year} {2009})}\BibitemShut {NoStop}%
\bibitem [{\citenamefont {Riggleman}\ \emph {et~al.}(2010)\citenamefont
  {Riggleman}, \citenamefont {Douglas},\ and\ \citenamefont
  {de~Pablo}}]{riggleman10b}%
  \BibitemOpen
  \bibfield  {author} {\bibinfo {author} {\bibfnamefont {R.~A.}\ \bibnamefont
  {Riggleman}}, \bibinfo {author} {\bibfnamefont {J.~F.}\ \bibnamefont
  {Douglas}}, \ and\ \bibinfo {author} {\bibfnamefont {J.~J.}\ \bibnamefont
  {de~Pablo}},\ }\bibfield  {title} {\enquote {\bibinfo {title}
  {Antiplasticization and the elastic properties of glass-forming polymer
  liquids},}\ }\href@noop {} {\bibfield  {journal} {\bibinfo  {journal} {Soft
  Matter}\ }\textbf {\bibinfo {volume} {6}},\ \bibinfo {pages} {292} (\bibinfo
  {year} {2010})}\BibitemShut {NoStop}%
\bibitem [{\citenamefont {Manning}\ and\ \citenamefont
  {Liu}(2011)}]{manning11}%
  \BibitemOpen
  \bibfield  {author} {\bibinfo {author} {\bibfnamefont {M.~L.}\ \bibnamefont
  {Manning}}\ and\ \bibinfo {author} {\bibfnamefont {A.~J.}\ \bibnamefont
  {Liu}},\ }\bibfield  {title} {\enquote {\bibinfo {title} {Vibrational modes
  identify soft spots in a sheared disordered packing},}\ }\href@noop {}
  {\bibfield  {journal} {\bibinfo  {journal} {Phys. Rev. Lett.}\ }\textbf
  {\bibinfo {volume} {107}},\ \bibinfo {pages} {108302} (\bibinfo {year}
  {2011})}\BibitemShut {NoStop}%
\bibitem [{\citenamefont {Rodney}\ and\ \citenamefont
  {Schroder}(2011)}]{rodney11}%
  \BibitemOpen
  \bibfield  {author} {\bibinfo {author} {\bibfnamefont {D.}~\bibnamefont
  {Rodney}}\ and\ \bibinfo {author} {\bibfnamefont {T.}~\bibnamefont
  {Schroder}},\ }\bibfield  {title} {\enquote {\bibinfo {title} {On the
  potential energy landscape of supercooled liquids and glasses},}\ }\href@noop
  {} {\bibfield  {journal} {\bibinfo  {journal} {Eur. Phys. J. E.}\ }\textbf
  {\bibinfo {volume} {34}},\ \bibinfo {pages} {100} (\bibinfo {year}
  {2011})}\BibitemShut {NoStop}%
\bibitem [{\citenamefont {Ding}\ \emph {et~al.}(2014)\citenamefont {Ding},
  \citenamefont {Patinet}, \citenamefont {Falk}, \citenamefont {Cheng},\ and\
  \citenamefont {Ma}}]{ding14}%
  \BibitemOpen
  \bibfield  {author} {\bibinfo {author} {\bibfnamefont {J.}~\bibnamefont
  {Ding}}, \bibinfo {author} {\bibfnamefont {S.}~\bibnamefont {Patinet}},
  \bibinfo {author} {\bibfnamefont {M.~L.}\ \bibnamefont {Falk}}, \bibinfo
  {author} {\bibfnamefont {Y.}~\bibnamefont {Cheng}}, \ and\ \bibinfo {author}
  {\bibfnamefont {E.}~\bibnamefont {Ma}},\ }\bibfield  {title} {\enquote
  {\bibinfo {title} {Soft spots and their structural signature in a metallic
  glass},}\ }\href@noop {} {\bibfield  {journal} {\bibinfo  {journal} {Proc.
  Nat. Acad. Sci.}\ }\textbf {\bibinfo {volume} {111}},\ \bibinfo {pages}
  {14052} (\bibinfo {year} {2014})}\BibitemShut {NoStop}%
\bibitem [{\citenamefont {Schoenholz}\ \emph {et~al.}(2014)\citenamefont
  {Schoenholz}, \citenamefont {Liu}, \citenamefont {Riggleman},\ and\
  \citenamefont {Rottler}}]{schoenholz14}%
  \BibitemOpen
  \bibfield  {author} {\bibinfo {author} {\bibfnamefont {S.~S.}\ \bibnamefont
  {Schoenholz}}, \bibinfo {author} {\bibfnamefont {A.~J.}\ \bibnamefont {Liu}},
  \bibinfo {author} {\bibfnamefont {R.~A.}\ \bibnamefont {Riggleman}}, \ and\
  \bibinfo {author} {\bibfnamefont {J.}~\bibnamefont {Rottler}},\ }\bibfield
  {title} {\enquote {\bibinfo {title} {Understanding plastic deformation in
  thermal glasses from single-soft-spot dynamics},}\ }\href@noop {} {\bibfield
  {journal} {\bibinfo  {journal} {Phys. Rev. X}\ }\textbf {\bibinfo {volume}
  {4}},\ \bibinfo {pages} {031014} (\bibinfo {year} {2014})}\BibitemShut
  {NoStop}%
\bibitem [{\citenamefont {Swayamjyoti}\ \emph {et~al.}(2014)\citenamefont
  {Swayamjyoti}, \citenamefont {L{\"o}ffler},\ and\ \citenamefont
  {Derlet}}]{swayamjyoti14}%
  \BibitemOpen
  \bibfield  {author} {\bibinfo {author} {\bibfnamefont {S.}~\bibnamefont
  {Swayamjyoti}}, \bibinfo {author} {\bibfnamefont {J.~F.}\ \bibnamefont
  {L{\"o}ffler}}, \ and\ \bibinfo {author} {\bibfnamefont {P.~M.}\ \bibnamefont
  {Derlet}},\ }\bibfield  {title} {\enquote {\bibinfo {title} {Local structural
  excitations in model glasses},}\ }\href@noop {} {\bibfield  {journal}
  {\bibinfo  {journal} {Phys. Rev. B}\ }\textbf {\bibinfo {volume} {89}},\
  \bibinfo {pages} {224201} (\bibinfo {year} {2014})}\BibitemShut {NoStop}%
\bibitem [{\citenamefont {Swayamjyoti}\ \emph {et~al.}(2016)\citenamefont
  {Swayamjyoti}, \citenamefont {L{\"o}ffler},\ and\ \citenamefont
  {Derlet}}]{swayamjyoti16}%
  \BibitemOpen
  \bibfield  {author} {\bibinfo {author} {\bibfnamefont {S.}~\bibnamefont
  {Swayamjyoti}}, \bibinfo {author} {\bibfnamefont {J.~F.}\ \bibnamefont
  {L{\"o}ffler}}, \ and\ \bibinfo {author} {\bibfnamefont {P.~M.}\ \bibnamefont
  {Derlet}},\ }\bibfield  {title} {\enquote {\bibinfo {title} {Local structural
  excitations in model glass systems under applied load},}\ }\href@noop {}
  {\bibfield  {journal} {\bibinfo  {journal} {Phys. Rev. B}\ }\textbf {\bibinfo
  {volume} {93}},\ \bibinfo {pages} {144202} (\bibinfo {year}
  {2016})}\BibitemShut {NoStop}%
\bibitem [{\citenamefont {Patinet}\ \emph {et~al.}(2016)\citenamefont
  {Patinet}, \citenamefont {Vandembroucq},\ and\ \citenamefont
  {Falk}}]{patinet16}%
  \BibitemOpen
  \bibfield  {author} {\bibinfo {author} {\bibfnamefont {S.}~\bibnamefont
  {Patinet}}, \bibinfo {author} {\bibfnamefont {D.}~\bibnamefont
  {Vandembroucq}}, \ and\ \bibinfo {author} {\bibfnamefont {M.~L.}\
  \bibnamefont {Falk}},\ }\bibfield  {title} {\enquote {\bibinfo {title}
  {Connecting local yield stresses with plastic activity in amorphous
  solids},}\ }\href@noop {} {\bibfield  {journal} {\bibinfo  {journal} {Phys.
  Rev. Lett.}\ }\textbf {\bibinfo {volume} {117}},\ \bibinfo {pages} {045501}
  (\bibinfo {year} {2016})}\BibitemShut {NoStop}%
\bibitem [{\citenamefont {Sollich}\ \emph {et~al.}(1997)\citenamefont
  {Sollich}, \citenamefont {Lequeux}, \citenamefont {Hebraud},\ and\
  \citenamefont {Cates}}]{sollich97}%
  \BibitemOpen
  \bibfield  {author} {\bibinfo {author} {\bibfnamefont {P.}~\bibnamefont
  {Sollich}}, \bibinfo {author} {\bibfnamefont {F.}~\bibnamefont {Lequeux}},
  \bibinfo {author} {\bibfnamefont {P.}~\bibnamefont {Hebraud}}, \ and\
  \bibinfo {author} {\bibfnamefont {M.~E.}\ \bibnamefont {Cates}},\ }\bibfield
  {title} {\enquote {\bibinfo {title} {Rheology of soft glassy materials},}\
  }\href@noop {} {\bibfield  {journal} {\bibinfo  {journal} {Phys. Rev. Lett.}\
  }\textbf {\bibinfo {volume} {78}},\ \bibinfo {pages} {2020} (\bibinfo {year}
  {1997})}\BibitemShut {NoStop}%
\bibitem [{\citenamefont {Sollich}(1998)}]{sollich98}%
  \BibitemOpen
  \bibfield  {author} {\bibinfo {author} {\bibfnamefont {P.}~\bibnamefont
  {Sollich}},\ }\bibfield  {title} {\enquote {\bibinfo {title} {Rheological
  constitutive equation for a model of soft glassy materials},}\ }\href@noop {}
  {\bibfield  {journal} {\bibinfo  {journal} {Phys. Rev. E}\ }\textbf {\bibinfo
  {volume} {58}},\ \bibinfo {pages} {738} (\bibinfo {year} {1998})}\BibitemShut
  {NoStop}%
\bibitem [{\citenamefont {Fielding}\ \emph {et~al.}(2000)\citenamefont
  {Fielding}, \citenamefont {Sollich},\ and\ \citenamefont
  {Cates}}]{fielding00}%
  \BibitemOpen
  \bibfield  {author} {\bibinfo {author} {\bibfnamefont {S.~M.}\ \bibnamefont
  {Fielding}}, \bibinfo {author} {\bibfnamefont {P.}~\bibnamefont {Sollich}}, \
  and\ \bibinfo {author} {\bibfnamefont {M.~E.}\ \bibnamefont {Cates}},\
  }\bibfield  {title} {\enquote {\bibinfo {title} {Aging and rheology in soft
  materials},}\ }\href@noop {} {\bibfield  {journal} {\bibinfo  {journal} {J.
  Rheol.}\ }\textbf {\bibinfo {volume} {44}},\ \bibinfo {pages} {323} (\bibinfo
  {year} {2000})}\BibitemShut {NoStop}%
\bibitem [{\citenamefont {Falk}\ and\ \citenamefont {Langer}(1998)}]{falk98}%
  \BibitemOpen
  \bibfield  {author} {\bibinfo {author} {\bibfnamefont {M.~L.}\ \bibnamefont
  {Falk}}\ and\ \bibinfo {author} {\bibfnamefont {J.~S.}\ \bibnamefont
  {Langer}},\ }\bibfield  {title} {\enquote {\bibinfo {title} {Dynamics of
  viscoplastic deformation in amorphous solids},}\ }\href@noop {} {\bibfield
  {journal} {\bibinfo  {journal} {Phys. Rev. E}\ }\textbf {\bibinfo {volume}
  {57}},\ \bibinfo {pages} {7192} (\bibinfo {year} {1998})}\BibitemShut
  {NoStop}%
\bibitem [{\citenamefont {Falk}\ \emph {et~al.}(2004)\citenamefont {Falk},
  \citenamefont {Langer},\ and\ \citenamefont {Pechenik}}]{falk04}%
  \BibitemOpen
  \bibfield  {author} {\bibinfo {author} {\bibfnamefont {M.~L.}\ \bibnamefont
  {Falk}}, \bibinfo {author} {\bibfnamefont {J.~S.}\ \bibnamefont {Langer}}, \
  and\ \bibinfo {author} {\bibfnamefont {L.}~\bibnamefont {Pechenik}},\
  }\bibfield  {title} {\enquote {\bibinfo {title} {Thermal effects in the
  shear-transformation-zone theory of amorphous plasticity: Comparisons to
  metallic glass data},}\ }\href@noop {} {\bibfield  {journal} {\bibinfo
  {journal} {Phys. Rev. E}\ }\textbf {\bibinfo {volume} {70}} ,\ \bibinfo
  {pages} {011507} (\bibinfo {year}
  {2004})}\BibitemShut {NoStop}%
\bibitem [{\citenamefont {Bouchbinder}\ and\ \citenamefont
  {Langer}(2009{\natexlab{a}})}]{bouchbinder09a}%
  \BibitemOpen
  \bibfield  {author} {\bibinfo {author} {\bibfnamefont {E.}~\bibnamefont
  {Bouchbinder}}\ and\ \bibinfo {author} {\bibfnamefont {J.~S.}\ \bibnamefont
  {Langer}},\ }\bibfield  {title} {\enquote {\bibinfo {title} {Nonequilibrium
  thermodynamics of driven amorphous materials. i. internal degrees of freedom
  and volume deformation},}\ }\href@noop {} {\bibfield  {journal} {\bibinfo
  {journal} {Phys. Rev. E}\ }\textbf {\bibinfo {volume} {80}},\ \bibinfo
  {pages} {031132} (\bibinfo {year} {2009}{\natexlab{a}})}\BibitemShut
  {NoStop}%
\bibitem [{\citenamefont {Bouchbinder}\ and\ \citenamefont
  {Langer}(2009{\natexlab{b}})}]{bouchbinder09b}%
  \BibitemOpen
  \bibfield  {author} {\bibinfo {author} {\bibfnamefont {E.}~\bibnamefont
  {Bouchbinder}}\ and\ \bibinfo {author} {\bibfnamefont {J.~S.}\ \bibnamefont
  {Langer}},\ }\bibfield  {title} {\enquote {\bibinfo {title} {Nonequilibrium
  thermodynamics of driven amorphous materials. ii. effective-temperature
  theory},}\ }\href@noop {} {\bibfield  {journal} {\bibinfo  {journal} {Phys.
  Rev. E}\ }\textbf {\bibinfo {volume} {80}},\ \bibinfo {pages} {031132}
  (\bibinfo {year} {2009}{\natexlab{b}})}\BibitemShut {NoStop}%
\bibitem [{\citenamefont {Langer}\ and\ \citenamefont
  {Egami}(2012)}]{langer12}%
  \BibitemOpen
  \bibfield  {author} {\bibinfo {author} {\bibfnamefont {J.~S.}\ \bibnamefont
  {Langer}}\ and\ \bibinfo {author} {\bibfnamefont {T.}~\bibnamefont {Egami}},\
  }\bibfield  {title} {\enquote {\bibinfo {title} {Glass dynamics at high
  strain rates},}\ }\href@noop {} {\bibfield  {journal} {\bibinfo  {journal}
  {Phys. Rev. E}\ }\textbf {\bibinfo {volume} {86}},\ \bibinfo {pages} {011502}
  (\bibinfo {year} {2012})}\BibitemShut {NoStop}%
\bibitem [{\citenamefont {Sollich}\ and\ \citenamefont
  {Cates}(2012)}]{sollich12}%
  \BibitemOpen
  \bibfield  {author} {\bibinfo {author} {\bibfnamefont {P.}~\bibnamefont
  {Sollich}}\ and\ \bibinfo {author} {\bibfnamefont {M.~E.}\ \bibnamefont
  {Cates}},\ }\bibfield  {title} {\enquote {\bibinfo {title} {Thermodynamic
  interpretation of soft glassy rheology models},}\ }\href@noop {} {\bibfield
  {journal} {\bibinfo  {journal} {Phys. Rev. E}\ }\textbf {\bibinfo {volume}
  {85}},\ \bibinfo {pages} {031127} (\bibinfo {year} {2012})}\BibitemShut
  {NoStop}%
\bibitem [{\citenamefont {Bouchbinder}\ and\ \citenamefont
  {Langer}(2013)}]{bouchbinder13}%
  \BibitemOpen
  \bibfield  {author} {\bibinfo {author} {\bibfnamefont {E.}~\bibnamefont
  {Bouchbinder}}\ and\ \bibinfo {author} {\bibfnamefont {J.~S.}\ \bibnamefont
  {Langer}},\ }\bibfield  {title} {\enquote {\bibinfo {title} {Nonequilibrium
  thermodynamics and glassy rheology},}\ }\href@noop {} {\bibfield  {journal}
  {\bibinfo  {journal} {Soft Matt.}\ }\textbf {\bibinfo {volume} {9}},\
  \bibinfo {pages} {8786} (\bibinfo {year} {2013})}\BibitemShut {NoStop}%
\bibitem [{\citenamefont {Fuereder}\ and\ \citenamefont
  {Ilg}(2013)}]{fuereder13}%
  \BibitemOpen
  \bibfield  {author} {\bibinfo {author} {\bibfnamefont {I.}~\bibnamefont
  {Fuereder}}\ and\ \bibinfo {author} {\bibfnamefont {P.}~\bibnamefont {Ilg}},\
  }\bibfield  {title} {\enquote {\bibinfo {title} {Nonequilibrium
  thermodynamics of the soft glassy rheology model},}\ }\href@noop {}
  {\bibfield  {journal} {\bibinfo  {journal} {Phys. Rev. E}\ }\textbf {\bibinfo
  {volume} {88}},\ \bibinfo {pages} {042134} (\bibinfo {year}
  {2013})}\BibitemShut {NoStop}%
\bibitem [{\citenamefont {Kamrin}\ and\ \citenamefont
  {Bouchbinder}(2014)}]{kamrin14}%
  \BibitemOpen
  \bibfield  {author} {\bibinfo {author} {\bibfnamefont {K.}~\bibnamefont
  {Kamrin}}\ and\ \bibinfo {author} {\bibfnamefont {E.}~\bibnamefont
  {Bouchbinder}},\ }\bibfield  {title} {\enquote {\bibinfo {title}
  {Two-temperature thermodynamics of deforming amorphous solids},}\ }\href@noop
  {} {\bibfield  {journal} {\bibinfo  {journal} {J. Mech. Phys. Solids}\
  }\textbf {\bibinfo {volume} {73}},\ \bibinfo {pages} {269} (\bibinfo {year}
  {2014})}\BibitemShut {NoStop}%
\bibitem [{\citenamefont {Bouchard}(1992)}]{bouchaud92}%
  \BibitemOpen
  \bibfield  {author} {\bibinfo {author} {\bibfnamefont {J.~P.}\ \bibnamefont
  {Bouchard}},\ }\bibfield  {title} {\enquote {\bibinfo {title} {Weak
  ergodicity breaking and aging in disordered systems},}\ }\href@noop {}
  {\bibfield  {journal} {\bibinfo  {journal} {J. Physique I}\ }\textbf
  {\bibinfo {volume} {2}},\ \bibinfo {pages} {1705} (\bibinfo {year}
  {1992})}\BibitemShut {NoStop}%
\bibitem [{\citenamefont {Monthus}\ and\ \citenamefont
  {Bouchard}(1996)}]{monthus96}%
  \BibitemOpen
  \bibfield  {author} {\bibinfo {author} {\bibfnamefont {C.}~\bibnamefont
  {Monthus}}\ and\ \bibinfo {author} {\bibfnamefont {J.~P.}\ \bibnamefont
  {Bouchard}},\ }\bibfield  {title} {\enquote {\bibinfo {title} {Models of
  traps and glass phenomenology},}\ }\href@noop {} {\bibfield  {journal}
  {\bibinfo  {journal} {J. Phys. A - Math. Gen.}\ }\textbf {\bibinfo {volume}
  {14}},\ \bibinfo {pages} {3847} (\bibinfo {year} {1996})}\BibitemShut
  {NoStop}%
\bibitem [{\citenamefont {Warren}\ and\ \citenamefont
  {Rottler}(2008)}]{warren08}%
  \BibitemOpen
  \bibfield  {author} {\bibinfo {author} {\bibfnamefont {M.}~\bibnamefont
  {Warren}}\ and\ \bibinfo {author} {\bibfnamefont {J.}~\bibnamefont
  {Rottler}},\ }\bibfield  {title} {\enquote {\bibinfo {title} {Mechanical
  rejuvenation and overaging in the soft glassy rheology model},}\ }\href@noop
  {} {\bibfield  {journal} {\bibinfo  {journal} {Phys. Rev. E}\ }\textbf
  {\bibinfo {volume} {78}},\ \bibinfo {pages} {041502} (\bibinfo {year}
  {2008})}\BibitemShut {NoStop}%
\bibitem [{\citenamefont {Fielding}(2014)}]{fielding14}%
  \BibitemOpen
  \bibfield  {author} {\bibinfo {author} {\bibfnamefont {S.~M.}\ \bibnamefont
  {Fielding}},\ }\bibfield  {title} {\enquote {\bibinfo {title} {Shear banding
  in soft glassy materials},}\ }\href@noop {} {\bibfield  {journal} {\bibinfo
  {journal} {Rep. Prog. Phys.}\ }\textbf {\bibinfo {volume} {77}},\ \bibinfo
  {pages} {102601} (\bibinfo {year} {2014})}\BibitemShut {NoStop}%
\bibitem [{\citenamefont {Merabia}\ and\ \citenamefont
  {Detcheverry}(2016)}]{merabia16}%
  \BibitemOpen
  \bibfield  {author} {\bibinfo {author} {\bibfnamefont {S.}~\bibnamefont
  {Merabia}}\ and\ \bibinfo {author} {\bibfnamefont {F.}~\bibnamefont
  {Detcheverry}},\ }\bibfield  {title} {\enquote {\bibinfo {title} {Thermally
  activated creep and fluidization in flowing disordered materials},}\
  }\href@noop {} {\bibfield  {journal} {\bibinfo  {journal} {Europhys. Lett.}\
  }\textbf {\bibinfo {volume} {116}},\ \bibinfo {pages} {46003} (\bibinfo
  {year} {2016})}\BibitemShut {NoStop}%
\bibitem [{\citenamefont {Radhakrishnan}\ and\ \citenamefont
  {Fielding}(2016)}]{radhakrishnan17}%
  \BibitemOpen
  \bibfield  {author} {\bibinfo {author} {\bibfnamefont {R.}~\bibnamefont
  {Radhakrishnan}}\ and\ \bibinfo {author} {\bibfnamefont {S.~M.}\ \bibnamefont
  {Fielding}},\ }\bibfield  {title} {\enquote {\bibinfo {title} {Shear banding
  of soft glassy materials in large amplitude oscillatory shear},}\ }\href@noop
  {} {\bibfield  {journal} {\bibinfo  {journal} {Phys. Rev. Lett.}\ }\textbf
  {\bibinfo {volume} {117}},\ \bibinfo {pages} {188001} (\bibinfo {year}
  {2016})}\BibitemShut {NoStop}%
\bibitem [{\citenamefont {Schuh}\ \emph {et~al.}(2007)\citenamefont {Schuh},
  \citenamefont {Hufnagel},\ and\ \citenamefont {Ramamurty}}]{schuh07}%
  \BibitemOpen
  \bibfield  {author} {\bibinfo {author} {\bibfnamefont {C.~A.}\ \bibnamefont
  {Schuh}}, \bibinfo {author} {\bibfnamefont {T.~C.}\ \bibnamefont {Hufnagel}},
  \ and\ \bibinfo {author} {\bibfnamefont {U.}~\bibnamefont {Ramamurty}},\
  }\bibfield  {title} {\enquote {\bibinfo {title} {Mechanical behavior of
  amorphous alloys},}\ }\href@noop {} {\bibfield  {journal} {\bibinfo
  {journal} {Acta Mat.}\ }\textbf {\bibinfo {volume} {55}},\ \bibinfo {pages}
  {4067} (\bibinfo {year} {2007})}\BibitemShut {NoStop}%
\bibitem [{\citenamefont {Roth}(2016)}]{roth16}%
  \BibitemOpen
  \bibinfo {editor} {\bibfnamefont {C.~B.}\ \bibnamefont {Roth}},\ ed.,\
  \href@noop {} {\emph {\bibinfo {title} {Polymer Glasses}}}\ (\bibinfo
  {publisher} {CRC Press},\ \bibinfo {year} {2016})\BibitemShut {NoStop}%
\bibitem [{\citenamefont {Bouchard}\ and\ \citenamefont
  {M{\'e}zard}(1997)}]{bouchaud97}%
  \BibitemOpen
  \bibfield  {author} {\bibinfo {author} {\bibfnamefont {J.~P.}\ \bibnamefont
  {Bouchard}}\ and\ \bibinfo {author} {\bibfnamefont {M.}~\bibnamefont
  {M{\'e}zard}},\ }\bibfield  {title} {\enquote {\bibinfo {title} {Universality
  classes for extreme-value statistics},}\ }\href@noop {} {\bibfield  {journal}
  {\bibinfo  {journal} {J. Phys. A: Math. Gen.}\ }\textbf {\bibinfo {volume}
  {30}},\ \bibinfo {pages} {7997} (\bibinfo {year} {1997})}\BibitemShut
  {NoStop}%  
\bibitem [{\citenamefont {Cates}\ and\ \citenamefont
  {Sollich}(2004)}]{cates04}%
  \BibitemOpen
  \bibfield  {author} {\bibinfo {author} {\bibfnamefont {M.~E.}\ \bibnamefont
  {Cates}}\ and\ \bibinfo {author} {\bibfnamefont {P.}~\bibnamefont
  {Sollich}},\ }\bibfield  {title} {\enquote {\bibinfo {title} {Tensorial
  constitutive models for disordered foams, dense emulsions, and other soft
  nonergodic materials},}\ }\href@noop {} {\bibfield  {journal} {\bibinfo
  {journal} {J. Rheo.}\ }\textbf {\bibinfo {volume} {48}},\ \bibinfo {pages}
  {193} (\bibinfo {year} {2004})}\BibitemShut {NoStop}%
\bibitem [{foo({\natexlab{a}})}]{footexp}%
  \BibitemOpen
  \href@noop {} {} \bibinfo {note} {Larger values of $\Delta
  t$ and hence $\Delta\epsilon = \dot\epsilon\Delta t$ can be employed if the
  [$1 - \Delta t/\tau(x_j,\epsilon_{j-1},T)$] term in Eq.\ \ref{eq:flow} is
  replaced by $\exp[-\Delta t/\tau(x_j,\epsilon_{j-1},T)]$ and the inflow term
  is modified accordingly. Here we do so.}\BibitemShut {Stop}%
\bibitem [{cod()}]{codelink}%
  \BibitemOpen
  \href@noop {} {}\bibinfo {note}
  {http://labs.cas.usf.edu/softmattertheory/}\BibitemShut {NoStop}%
\bibitem [{\citenamefont {Johnson}\ \emph {et~al.}(2002)\citenamefont
  {Johnson}, \citenamefont {Lu},\ and\ \citenamefont {Demetriou}}]{johnson02}%
  \BibitemOpen
  \bibfield  {author} {\bibinfo {author} {\bibfnamefont {W.~L.}\ \bibnamefont
  {Johnson}}, \bibinfo {author} {\bibfnamefont {J.}~\bibnamefont {Lu}}, \ and\
  \bibinfo {author} {\bibfnamefont {M.~D.}\ \bibnamefont {Demetriou}},\
  }\bibfield  {title} {\enquote {\bibinfo {title} {Deformation and flow in bulk
  metallic glasses and deeply undercooled glass forming liquids: a self
  consistent dynamic free volume model},}\ }\href@noop {} {\bibfield  {journal}
  {\bibinfo  {journal} {Intermetallics}\ }\textbf {\bibinfo {volume} {10}},\
  \bibinfo {pages} {1039} (\bibinfo {year} {2002})}\BibitemShut {NoStop}%
\bibitem [{\citenamefont {Lu}\ \emph {et~al.}(2003)\citenamefont {Lu},
  \citenamefont {Ravichandran},\ and\ \citenamefont {Johnson}}]{lu03}%
  \BibitemOpen
  \bibfield  {author} {\bibinfo {author} {\bibfnamefont {J.}~\bibnamefont
  {Lu}}, \bibinfo {author} {\bibfnamefont {G.}~\bibnamefont {Ravichandran}}, \
  and\ \bibinfo {author} {\bibfnamefont {W.~L.}\ \bibnamefont {Johnson}},\
  }\bibfield  {title} {\enquote {\bibinfo {title} {Deformation behavior of the
  $\rm{Zr_{41.2}Ti_{13.8}Cu_{12.5}Ni_{10}Be_{22.5}}$ bulk metallic glass over a wide
  range of strain-rates and temperatures},}\ }\href@noop {} {\bibfield
  {journal} {\bibinfo  {journal} {Acta. Mat.}\ }\textbf {\bibinfo {volume}
  {51}},\ \bibinfo {pages} {3429} (\bibinfo {year} {2003})}\BibitemShut
  {NoStop}%
\bibitem [{\citenamefont {Lee}\ \emph {et~al.}(2008)\citenamefont {Lee},
  \citenamefont {Paeng}, \citenamefont {Swallen},\ and\ \citenamefont
  {Ediger}}]{lee09}%
  \BibitemOpen
  \bibfield  {author} {\bibinfo {author} {\bibfnamefont {H.-N.}\ \bibnamefont
  {Lee}}, \bibinfo {author} {\bibfnamefont {K.}~\bibnamefont {Paeng}}, \bibinfo
  {author} {\bibfnamefont {S.~F.}\ \bibnamefont {Swallen}}, \ and\ \bibinfo
  {author} {\bibfnamefont {M.~D.}\ \bibnamefont {Ediger}},\ }\bibfield  {title}
  {\enquote {\bibinfo {title} {Direct measurement of molecular mobility in
  actively deformed polymer glasses},}\ }\href@noop {} {\bibfield  {journal}
  {\bibinfo  {journal} {Science}\ }\textbf {\bibinfo {volume} {323}},\ \bibinfo
  {pages} {232--234} (\bibinfo {year} {2008})}\BibitemShut {NoStop}%
\bibitem [{\citenamefont {Bending}\ \emph {et~al.}(2014)\citenamefont
  {Bending}, \citenamefont {Christison}, \citenamefont {Ricci},\ and\
  \citenamefont {Ediger}}]{bending14}%
  \BibitemOpen
  \bibfield  {author} {\bibinfo {author} {\bibfnamefont {B.}~\bibnamefont
  {Bending}}, \bibinfo {author} {\bibfnamefont {K.}~\bibnamefont {Christison}},
  \bibinfo {author} {\bibfnamefont {J.}~\bibnamefont {Ricci}}, \ and\ \bibinfo
  {author} {\bibfnamefont {M.~D.}\ \bibnamefont {Ediger}},\ }\bibfield  {title}
  {\enquote {\bibinfo {title} {Measurement of segmental mobility during
  constant strain rate deformation of a poly(methyl methacrylate) glass},}\
  }\href@noop {} {\bibfield  {journal} {\bibinfo  {journal} {Macromolecules}\
  }\textbf {\bibinfo {volume} {47}},\ \bibinfo {pages} {800--806} (\bibinfo
  {year} {2014})}\BibitemShut {NoStop}%
\bibitem [{\citenamefont {Bending}\ and\ \citenamefont
  {Ediger}(2016)}]{bending16}%
  \BibitemOpen
  \bibfield  {author} {\bibinfo {author} {\bibfnamefont {B.}~\bibnamefont
  {Bending}}\ and\ \bibinfo {author} {\bibfnamefont {M.~D.}\ \bibnamefont
  {Ediger}},\ }\bibfield  {title} {\enquote {\bibinfo {title} {Comparison of
  mechanical and molecular measures of mobility during constant strain rate
  deformation of a pmma glass},}\ }\href@noop {} {\bibfield  {journal}
  {\bibinfo  {journal} {J. Pol. Sci. Part B: Pol. Phys.}\ }\textbf {\bibinfo
  {volume} {54}},\ \bibinfo {pages} {1957} (\bibinfo {year}
  {2016})}\BibitemShut {NoStop}%
\bibitem [{\citenamefont {Dequidt}\ \emph {et~al.}(2016)\citenamefont
  {Dequidt}, \citenamefont {Conca}, \citenamefont {Delannoy}, \citenamefont
  {Sotta}, \citenamefont {Lequeux},\ and\ \citenamefont {Long}}]{dequidt16}%
  \BibitemOpen
  \bibfield  {author} {\bibinfo {author} {\bibfnamefont {A.}~\bibnamefont
  {Dequidt}}, \bibinfo {author} {\bibfnamefont {L.}~\bibnamefont {Conca}},
  \bibinfo {author} {\bibfnamefont {{J.-Y.}}\ \bibnamefont {Delannoy}},
  \bibinfo {author} {\bibfnamefont {P.}~\bibnamefont {Sotta}}, \bibinfo
  {author} {\bibfnamefont {F.}~\bibnamefont {Lequeux}}, \ and\ \bibinfo
  {author} {\bibfnamefont {D.~R.}\ \bibnamefont {Long}},\ }\bibfield  {title}
  {\enquote {\bibinfo {title} {Heterogeneous dynamics and polymer
  plasticity},}\ }\href@noop {} {\bibfield  {journal} {\bibinfo  {journal}
  {Macromolecules}\ }\textbf {\bibinfo {volume} {49}},\ \bibinfo {pages} {9148}
  (\bibinfo {year} {2016})}\BibitemShut {NoStop}%
\bibitem [{\citenamefont {Hebraud}\ and\ \citenamefont
  {Lequeux}(1998)}]{hebraud98}%
  \BibitemOpen
  \bibfield  {author} {\bibinfo {author} {\bibfnamefont {P.}~\bibnamefont
  {Hebraud}}\ and\ \bibinfo {author} {\bibfnamefont {F.}~\bibnamefont
  {Lequeux}},\ }\bibfield  {title} {\enquote {\bibinfo {title} {Mode-coupling
  theory for the pasty rheology of soft glassy materials},}\ }\href@noop {}
  {\bibfield  {journal} {\bibinfo  {journal} {Phys. Rev. Lett.}\ }\textbf
  {\bibinfo {volume} {81}},\ \bibinfo {pages} {2934} (\bibinfo {year}
  {1998})}\BibitemShut {NoStop}%    
\bibitem [{\citenamefont {Arruda}\ \emph {et~al.}(1995)\citenamefont {Arruda},
  \citenamefont {Boyce},\ and\ \citenamefont {Jayachandran}}]{arruda95}%
  \BibitemOpen
  \bibfield  {author} {\bibinfo {author} {\bibfnamefont {E.~M.}\ \bibnamefont
  {Arruda}}, \bibinfo {author} {\bibfnamefont {M.~C.}\ \bibnamefont {Boyce}}, \
  and\ \bibinfo {author} {\bibfnamefont {R.}~\bibnamefont {Jayachandran}},\
  }\bibfield  {title} {\enquote {\bibinfo {title} {Effects of strain rate,
  temperature and thermomechanical coupling on the finite strain deformation of
  glassy polymers},}\ }\href@noop {} {\bibfield  {journal} {\bibinfo  {journal}
  {Mech. Mat.}\ }\textbf {\bibinfo {volume} {19}},\ \bibinfo {pages} {193}
  (\bibinfo {year} {1995})}\BibitemShut {NoStop}%
\bibitem [{foo({\natexlab{b}})}]{footvias}%
  \BibitemOpen
  \href@noop {} {} \bibinfo {note} {An analogous approach
  was used to study the trap model's aging dynamics in Ref.\
  \cite{viasnoff02}.}\BibitemShut {Stop}%
\bibitem [{\citenamefont {Viasnoff}\ and\ \citenamefont
  {Lequeux}(2002)}]{viasnoff02}%
  \BibitemOpen
  \bibfield  {author} {\bibinfo {author} {\bibfnamefont {V.}~\bibnamefont
  {Viasnoff}}\ and\ \bibinfo {author} {\bibfnamefont {F.}~\bibnamefont
  {Lequeux}},\ }\bibfield  {title} {\enquote {\bibinfo {title} {Rejuvenation
  and overaging in a colloidal glass under shear},}\ }\href@noop {} {\bibfield
  {journal} {\bibinfo  {journal} {Phys. Rev. Lett.}\ }\textbf {\bibinfo
  {volume} {89}},\ \bibinfo {pages} {065701} (\bibinfo {year}
  {2002})}\BibitemShut {NoStop}%
\bibitem [{foo({\natexlab{c}})}]{footEgt0}%
  \BibitemOpen
  \href@noop {} {} \bibinfo {note} {Systems with $E(\epsilon) > 0$ are
  effectively ``unbound'' and would presumably fracture if the theoretical
  model allowed for fracture.}\BibitemShut {Stop}%
\bibitem [{\citenamefont {Robertson}(1966)}]{robertson66}%
  \BibitemOpen
  \bibfield  {author} {\bibinfo {author} {\bibfnamefont {R.~E.}\ \bibnamefont
  {Robertson}},\ }\bibfield  {title} {\enquote {\bibinfo {title} {Theory for
  the plasticity of glassy polymers},}\ }\href@noop {} {\bibfield  {journal}
  {\bibinfo  {journal} {J. Chem. Phys.}\ }\textbf {\bibinfo {volume} {44}},\
  \bibinfo {pages} {3950} (\bibinfo {year} {1966})}\BibitemShut {NoStop}%
\bibitem [{\citenamefont {Cugliandolo}\ \emph {et~al.}(1997)\citenamefont
  {Cugliandolo}, \citenamefont {Kurchan},\ and\ \citenamefont
  {Peliti}}]{cugliandolo97}%
  \BibitemOpen
  \bibfield  {author} {\bibinfo {author} {\bibfnamefont {L.~F.}\ \bibnamefont
  {Cugliandolo}}, \bibinfo {author} {\bibfnamefont {J.}~\bibnamefont
  {Kurchan}}, \ and\ \bibinfo {author} {\bibfnamefont {L.}~\bibnamefont
  {Peliti}},\ }\bibfield  {title} {\enquote {\bibinfo {title} {Energy flow,
  partial equilibration, and effective temperatures in systems with slow
  dynamics},}\ }\href@noop {} {\bibfield  {journal} {\bibinfo  {journal} {Phys.
  Rev. E}\ }\textbf {\bibinfo {volume} {55}},\ \bibinfo {pages} {3898}
  (\bibinfo {year} {1997})}\BibitemShut {NoStop}%
\bibitem [{\citenamefont {Berthier}\ \emph {et~al.}(2000)\citenamefont
  {Berthier}, \citenamefont {Barrat},\ and\ \citenamefont
  {Kurchan}}]{berthier00}%
  \BibitemOpen
  \bibfield  {author} {\bibinfo {author} {\bibfnamefont {L.}~\bibnamefont
  {Berthier}}, \bibinfo {author} {\bibfnamefont {{J.-L.}}\ \bibnamefont
  {Barrat}}, \ and\ \bibinfo {author} {\bibfnamefont {J.}~\bibnamefont
  {Kurchan}},\ }\bibfield  {title} {\enquote {\bibinfo {title} {A
  two-time-scale, two-temperature scenario for nonlinear rheology},}\
  }\href@noop {} {\bibfield  {journal} {\bibinfo  {journal} {Phys. Rev. E}\
  }\textbf {\bibinfo {volume} {61}},\ \bibinfo {pages} {5464} (\bibinfo {year}
  {2000})}\BibitemShut {NoStop}%
\end{thebibliography}
%\end{document}

%

\end{document}